\documentclass[preprint,12pt,authoryear]{elsarticle}

\usepackage{amssymb}
\usepackage{amsmath}
\usepackage{tikz}
	\usetikzlibrary{shapes,arrows, chains, positioning, shadings, calc, decorations}

\usepackage{natbib}
\journal{Results in Earth Sciences - Applied Artificial Intelligence (AI) and Machine Learning (ML) in Geosciences}

\begin{document}

\begin{frontmatter}



\title{Detecting Spatiotemporal $b$-Value Anomalies with a Progressive Deep Learning Architecture} 

\author[label1,label2]{Jonas Köhler} 
\author[label1]{Wei Li} 
\author[label1,label3]{Johannes Faber} 
\author[label1,label2]{Georg Rümpker} 
\author[label1,label2]{Nishtha Srivastava} 
\ead{N.Srivastava@em.uni-frankfurt.de}

\affiliation[label1]{organization={Frankfurt Institute of Advanced Studies},
            addressline={Ruth-Moufang-Str.~1},
            city={Frankfurt am Main},
            postcode={60438},
            state={Hessen},
            country={Germany}}

\affiliation[label2]{organization={Institute of Geosciences, Goethe-University Frankfurt},
            addressline={Altenhöferallee 1},
            city={Frankfurt am Main},
            postcode={60438},
            state={Hessen},
            country={Germany}}

\affiliation[label3]{organization={Institute for Theoretical Physics, Goethe Universität},
            addressline={Max-von-Laue-Str. 1},
            city={Frankfurt am Main},
            postcode={60438},
            state={Hessen},
            country={Germany}}


\newcommand{\MJMA}{$\mathrm{M_{JMA}}$}
\newcommand{\ML}{$\mathrm{M_{L}}$}
\newcommand{\MV}{$\mathrm{M_{V}}$}
\newcommand{\MS}{$\mathrm{M_{S}}$}
\newcommand{\mb}{$\mathrm{m_{b}}$}
\newcommand{\Mw}{$\mathrm{M_{W}}$}
\newcommand{\EQ}{\texttt{EQ}\ }
\newcommand{\nEQ}{\texttt{nEQ}\ }
\newcommand{\MBest}{\texttt{Model4.0}\ }
\newcommand{\MPerf}{\texttt{Model4.9}\ }
\newcommand{\thk}{Tōhoku\ }

\begin{abstract}
Identifying systematic patterns in seismicity that precede large earthquakes remains a central challenge in statistical seismology. In this work, we present a methodological framework for detecting spatiotemporal anomalies in seismicity using the evolution of gridded $b$-values. Focusing on the Japanese subduction zone, we construct daily $b$-value fields on a fine spatial grid by aggregating local seismicity over moving time windows, yielding a continuous 2+1D representation of seismic-state evolution.

We formulate the problem as a binary classification task in which spatiotemporal blocks extracted from these $b$-value fields are labeled according to the occurrence of a target earthquake with \Mw $\geq 5$ in the central region within the next day. To model this data, we introduce a hybrid deep-learning architecture that combines a spatial convolutional encoder with a temporal convolutional network, enabling joint learning of spatial structure and temporal dynamics. A progressive meta-epoch training scheme is employed, in which the model is iteratively updated using a time-forward strategy that mirrors operational deployment and mitigates issues related to nonstationarity.

This paper is strictly methodological in scope. It describes the construction of $b$-value fields, the spatiotemporal sampling strategy, the network architecture, and the progressive training and internal validation framework used for model development and parameter selection. 
\end{abstract}

\begin{graphicalabstract}
\includegraphics{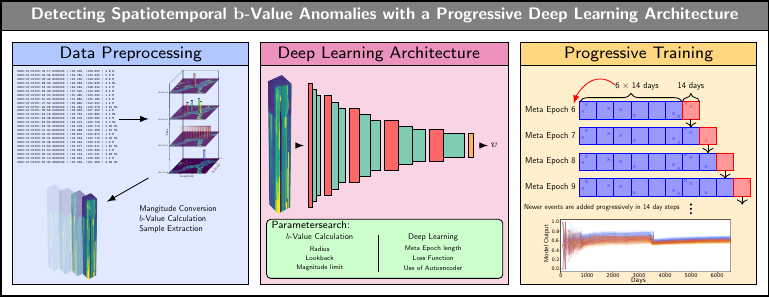}
\end{graphicalabstract}

\begin{highlights}
\item We develop an anomaly detection pipeline using spatiotemporal $b$-value variations
\item We perform a parameter search exploring possible parametric and methodlogical choices 
\item We employ a progressive training technique based on continuous training set updates to mimic real application
\end{highlights}

\begin{keyword}
deep learning \sep anomaly detection \sep b-value \sep progressive learning



\end{keyword}

\end{frontmatter}

\newcommand{\MJMA}{$\mathrm{M_{JMA}}$}
\newcommand{\ML}{$\mathrm{M_{L}}$}
\newcommand{\MV}{$\mathrm{M_{V}}$}
\newcommand{\MS}{$\mathrm{M_{S}}$}
\newcommand{\mb}{$\mathrm{m_{b}}$}
\newcommand{\Mw}{$\mathrm{M_{W}}$}
\newcommand{\EQ}{\texttt{EQ}\ }
\newcommand{\nEQ}{\texttt{nEQ}\ }
\newcommand{\MBest}{\texttt{Model4.0}\ }
\newcommand{\MPerf}{\texttt{Model4.9}\ }
\newcommand{\thk}{Tōhoku\ }

\newcommand{\bv}{$b$-value\ }
\newcommand{\bvs}{$b$-values\ }

%
%

\section{Introduction}
Earthquakes are among the most damaging natural hazards, and understanding the spatiotemporal organization of seismicity remains an open challenge in seismology. Decades of extensive data collection, combined with improved characterizations of seismic processes, have enabled increasingly detailed analyses of earthquake occurrence and its underlying physical drivers.

A substantial body of prior work has examined empirical regularities in
seismicity, most prominently the Gutenberg-Richter law \citep{Gutenberg1944},
\begin{equation}
    \log_{10} N = a - b M
\end{equation} which gives a logarithmic relation between the number of earthquakes $N$ above a certain magnitude $M$. The \bv gives an estimate for the occurrence rate of large earthquakes compared to smaller earthquakes. Spatial and temporal variations in the \bv have long been interpreted as indicators of stress conditions and have therefore been used in a variety of seismological applications, including the study of precursor-like anomalies \citep{Smith1981, Main1989}, the construction of earthquake rate models \citep{Smyth2011}, and the discrimination between foreshocks and mainshocks \citep{Gulia2019}. Statistical point-process methods such as the Epidemic Type Aftershock Sequence (ETAS) model \citep{Kagan1981, Kagan1987, Ogata1988, Ogata1998} further represent a widely used framework for characterizing the temporal organization of seismic activity and its aftershock triggering
dynamics.

In parallel with these developments, the rapid growth of Deep Learning has enabled substantial progress in processing both seismic waveforms and geodetic data \citep{LeCun2015}. Applications include event detection and magnitude estimation from seismograms \citep{Chakraborty2022}, deformation analysis using GNSS observations \citep{Quinteros2023}, seismic phase picking \citep{Zhu2018, Mousavi2020b, Li2022, Saad2023b}, and synthetic waveform generation \citep{lehmann2023}, among others \citep{Mousavi2022}. Recent work has also begun to explore the use of machine learning to extract patterns in seismicity catalogs. Examples range from models using multiple physical observables in spatiotemporal bins \citep{Fox2022} to those incorporating electromagnetic or geo-acoustic signals \citep{Saad2023}. Neural point-process approaches have similarly been applied to catalog data \citep{Stockmann2023}.

The combination of spatiotemporal \bv maps with Deep Learning was first introduced in \citep{Koehler2023}, followed by architectures that integrate statistical insights from the ETAS model \citep{Zhan2024} or operate on energy-release maps \citep{Zhan2025}.

A persistent difficulty for data-driven analyses of seismicity is that both the available sample size and the underlying data distribution evolve in time. This poses challenges for training neural networks on sequential data while avoiding information leakage from the future into earlier training periods. To address related issues in machine learning, progressive or curriculum-based training strategies have been developed \citep{Bengio2009}, in which a model is trained on data introduced in a structured temporal or difficulty-based sequence. Surveys of such approaches emphasize their utility for stabilizing learning under non-stationary conditions \citep{Soviany2021}, and progressive variants of neural architectures have been proposed to incrementally incorporate new information \citep{Siddiqui2021}. Despite their use in machine learning, these strategies have seen little adoption in seismological applications.

In this work, we focus on the methodological development of a framework that identifies spatiotemporal patterns in daily \bvs computed across Japan. The approach centers on a hybrid deep neural network that operates on $512 \times 32 \times 32$ blocks of \bvs and is trained through a progressive, time-forward meta-epoch scheme inspired by curriculum-learning principles. This design ensures that each temporal increment is used both as a validation window and as newly added training data, while preventing leakage of future information. Japan provides an ideal study region due to its high seismicity rate \citep{Wakita2013} and well-characterized catalog completeness \citep{Nanjo2010}. Tectonically, the region is shaped primarily by the subduction of the Pacific Plate beneath the Okhotsk Plate and the interaction of the Philippine and Eurasian Plates \citep{Bird2003} (though we focus on the first two). Throughout this paper, our objective is to present and analyze the methodological components of this framework, with an emphasis on the construction of spatiotemporal \bv fields, the design of the neural network architecture, and the progressive training strategy.\\

This paper presents a methodological framework for identifying spatiotemporal patterns in evolving \bvs. Our contributions are threefold.  \\
(1) We construct daily \bv fields on a fine spatial grid across Japan and assemble them into fixed-size spatiotemporal blocks suitable for neural processing.  \\
(2) We develop a hybrid convolutional-temporal architecture designed to extract localized spatial structure and longer-range temporal dependencies within these blocks.  \\
(3) We introduce a progressive meta-epoch training scheme that incrementally incorporates new data in a strictly time-forward manner, enabling stable model development under non-stationary catalog conditions while avoiding information leakage.  

\section{Data}
\begin{figure}
    \includegraphics[width=\linewidth]{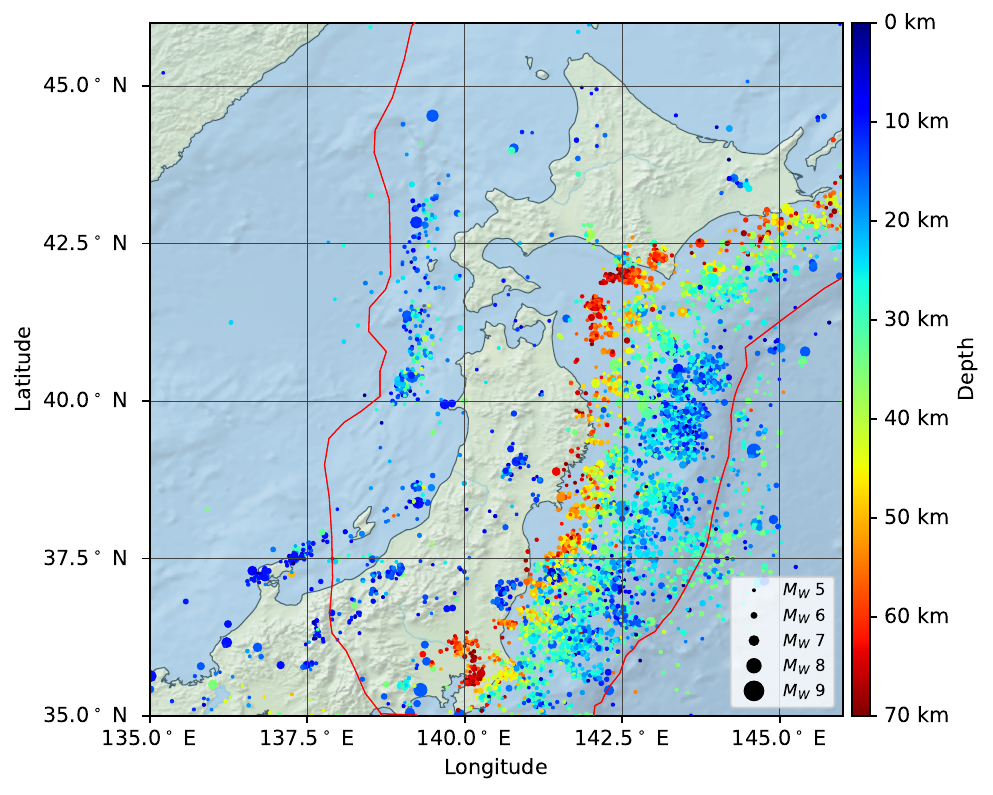}
    \caption{Overview of the study area. Shown are all earthquakes between 1999-01-01 to 2019-12-31 colored by depth and scaled by magnitude. The tectonic boundaries (red) are taken from \citep{Bird2003}.}
    \label{fig:Japan_Overview}
\end{figure}

For this study we focus our attention on the region $[35^\circ, 46^\circ \mathrm{N}] \times [135^\circ, 146^\circ \mathrm{E}]$ around Japan (see Figure \ref{fig:Japan_Overview}). This subarea is chosen to simplify the system: It reduces the relevant plate boundaries from three to one and limits the data to an area where most larger earthquakes occur within the uppermost 70~km.

We use the ISC Catalog \citep{ISC_1of3, ISC_2of3, ISC_3of3} from 1999-01-01 to 2019-12-31 to create our dataset (accessed date is September 10th, 2024), as for our area of interest the ISC contains more earthquakes than the USGS catalog. Additionally, we use the data from 2020-01-01 to 2022-12-31 as a test set in some instances.

For our date and depth range, the catalog consists of 1,393,018 events for the training and validation set and 347,674 for the test set. We convert the different magnitude types (\MS, \mb, \MV, \ML, \MJMA) to moment magnitude \Mw, which is otherwise not widely present in the catalog for smaller magnitudes, using the relations given in \citep{Scordilis2005}. 
\begin{align}
  M_w &=
    \begin{cases}
      0.65 M_s + 2.20 & 3.0 \leq M_s \leq 6.1\\
      1.00 M_s - 0.02 & 6.2 \leq M_s \leq 8.0\\
    \end{cases}\\
    M_w &= 0.85 m_b + 1.02 \\ 
    M_w &=
    \begin{cases}
      0.58 M_{JMA} + 2.25 & 3.0 \leq M_{JMA} \leq 5.5\\
      0.97 M_{JMA} + 0.04 & 5.6 \leq M_{JMA} \leq 8.2\\
    \end{cases}
\end{align}

If multiple magnitude types are available for a given event, we prioritize them according to the order specified in the bracket above. By restricting our analysis to only these magnitude types, we exclude only $\approx 2000$ events of smaller magnitudes ($\leq 3.5$) across the entire region.

Additionally, we limit earthquake depths to 70~km or less for two main reasons. First, catalog completeness varies with depth, and maintaining a consistent depth threshold ensures better homogeneity in the dataset. Second, the \bv is empirically known to vary with depth \cite{Mori1997}, a factor we do not explicitly consider. Constraining depth reduces potential errors introduced by ignoring depth variations in \bv calculations, thereby minimizing the influence of deeper, stronger earthquakes on surface-level \bv estimates.

The dataset includes the 2011 Tōhoku Earthquake, a megathrust event that significantly altered the seismicity distribution in the region of interest for several years. Although this inclusion may adversely affect the results, it is unavoidable because the catalog prior to 1999 lacks the same good completeness, and the post-2011 period is too brief and still dominated by elevated seismicity due to aftershocks from the Tōhoku event.

\subsection{$b$-Value Calculation}
Using the converted \Mw catalog we calculate the \bv on a fine $0.1^\circ$ by $0.1^\circ$ grid for each day after 2000-01-01. For each of these boxes, the \bv is calculated using the relation from Aki \citep{Aki1965}:

\begin{align}
b^\prime &= \frac{1}{\frac{\sum\limits_{i=1}^{n} M_i}{n} - M_0}\\
b &= \log_{10}(e) b^\prime 
\end{align}

To address catalog changing completeness magnitudes, we employ the $b$-positive method \citep{Elst2021} to select earthquakes used in the calculation. According to this method, only the positive differences in magnitudes between subsequent earthquakes are used for \bv calculation.

Earthquakes occurring within a radius $r$ of the center of the location and within the last $t$ days, akin to a spatiotemporal cylindrical stencil, are used for a single \bv calculation. The values for $r$ and $t$ are determined by a parameter search, which is detailed in the appendix (section \ref{ssec:parametersearch}).

This approach yields a \bv array of size $7305 \times 110 \times 110$. If fewer than two earthquakes are recorded, the \bv is set to 0. Similarly, the \bv is set to 0 when all magnitudes within a cylinder have the same value, which can occur if there are few recorded earthquakes. To prevent extreme values, the \bv is clipped between 0 and 2, as cases with only three earthquakes can result in unrealistically high \bvs. 

Although this method occasionally produces unrealistic \bv results in border regions of the seismic station coverage where cells with $b=0$ and $b=2$ frequently adjoin, it has been empirically validated and works well for this study. We do not take any special care of the catalog completeness besides the $b$-positive method, however a completeness analysis using the method from \citep{Godano2023} is shown in the Appendix (Figure~\ref{fig:completeness}).

\subsection{Labeled Sample Construction}
\label{sec:NEQdefinitionsection}
The creation of the dataset represents an important step for a machine learning approach because it embeds the information and format that the network will learn.
For our classification problem, we define two classes, each referring to a time series of spatial \bvs:
\vspace{-0.8em}
\begin{itemize}
    \item[(1)] an earthquake class (referred to as \EQ), defined by an earthquake of magnitude \Mw $\geq 5$ on the day after the series ends
    \item[(2)] an \nEQ class, defined by the absence of such an event:
    \begin{itemize}
        \item No earthquake with \Mw $ \geq 4.9$ within a $0.8^\circ$ L1 radius
        \item No earthquake with \Mw $ \geq 4.9$ within $\pm 7$ days
        \item An average of at least 10 earthquakes for \bv calculation
    \end{itemize}
\end{itemize}

The limit of 4.9 was chosen despite the parameter search.\footnote{Smaller values lead to a less representative \nEQ dataset, which will cause poor performance in case of a realistic forecasting scenario. The results of the parametrically optimal threshold of 4.0 will a also be discussed in \ref{sec:MBest}.}
The L1 radius refers to the taxicab geometry. The last criterion for \nEQ is necessary, so the model is not trained on that part of the region with too little or no data. While 10 earthquakes is usually to low to calculate a stable b value, increasing this further will remove too many samples from the data.

We exclude locations where there is an earthquake 7 days before and after, so that the model is not punished for getting the forecast of large earthquakes wrong by only a short amount of time.

As our \bv block starts on 2000-01-01 and we require 512 days of history for our classification later, the analysis is constrained to the events after 2001-05-27 (512 days for a history) and 2019-12-31. Within this time-frame there are 2328 \Mw $\geq 5$ events, and 1289 of those are unique in their spatiotemporal bin and far enough away from the domain borders (16 cells or $1.6^\circ$) to be considered. For each of those events, we carve a $512 \times 32 \times 32$ shaped sample centered around the epicenter of the event out of the \bv block, so that the earthquake would happen on the 513th day, exactly one day after the sample block ends.  The $[32 \times 32]$ corresponds to a $3.2^\circ \times 3.2^\circ$ area. The hypothesis here is that there are characteristic patterns in the \bv prior to a large event \citep{Main1989, Gulia2019}, and the network is tasked with finding them. 
On the other side of the 1289 larger events, we also need a class that constitutes ``calm'' situations in which spatiotemporal sequences are not followed by a larger event.

\section{Methods}
This section describes the formulation of the classification problem, the neural architecture, and the progressive, time-forward training protocol used to respect causality and to reduce concept drift, which could be caused by larger events or by an otherwise shifting seismic background.

\subsection{Problem formulation}
We cast the task as a binary classification: given a 512-day spatiotemporal history of \bv fields on a $32 \times 32$ spatial patch, predict whether a large earthquake will occur in the central location on the day immediately following the end of the input window. Positive samples (EQ) correspond to occurrences of \Mw $\geq 5$ at the central cell on day $t_\mathrm{end}+1$. Negative samples (\nEQ) correspond to the absence of a nearby large event within the spatiotemporal exclusion window defined in Section~\ref{sec:NEQdefinitionsection}. For clarity: the classifier output is an uncalibrated anomaly score (probability-like), not a direct rate forecast.

\subsection{Model Architecture}
\label{sec:architecture}
The classifier is designed to process spatiotemporal \bv fields represented as $512\times 32\times 32$ blocks. Depending on the configuration, the classifier receives either the raw \bv block or the reconstruction error produced by the optional autoencoder (described in  \ref{sec:AE}).

The architecture combines spatial 2D encoding with temporal convolutions inspired by temporal convolutional networks.  Spatial downsampling is implemented using 3D convolutions with kernel size $(1,2,2)$ and stride $(1,2,2)$, halving the spatial extent at each spatial stage while doubling the channel count. Temporal modeling uses 3D convolutions with kernel size $(2,1,1)$ and exponentially increasing dilation factors $(2^n,1,1)$ along the temporal axis. These dilated  convolutions give the model a receptive field that spans the full lookback period with minimal layer count and kernel size.

\begin{figure}
    \includegraphics[height=0.9\textheight]{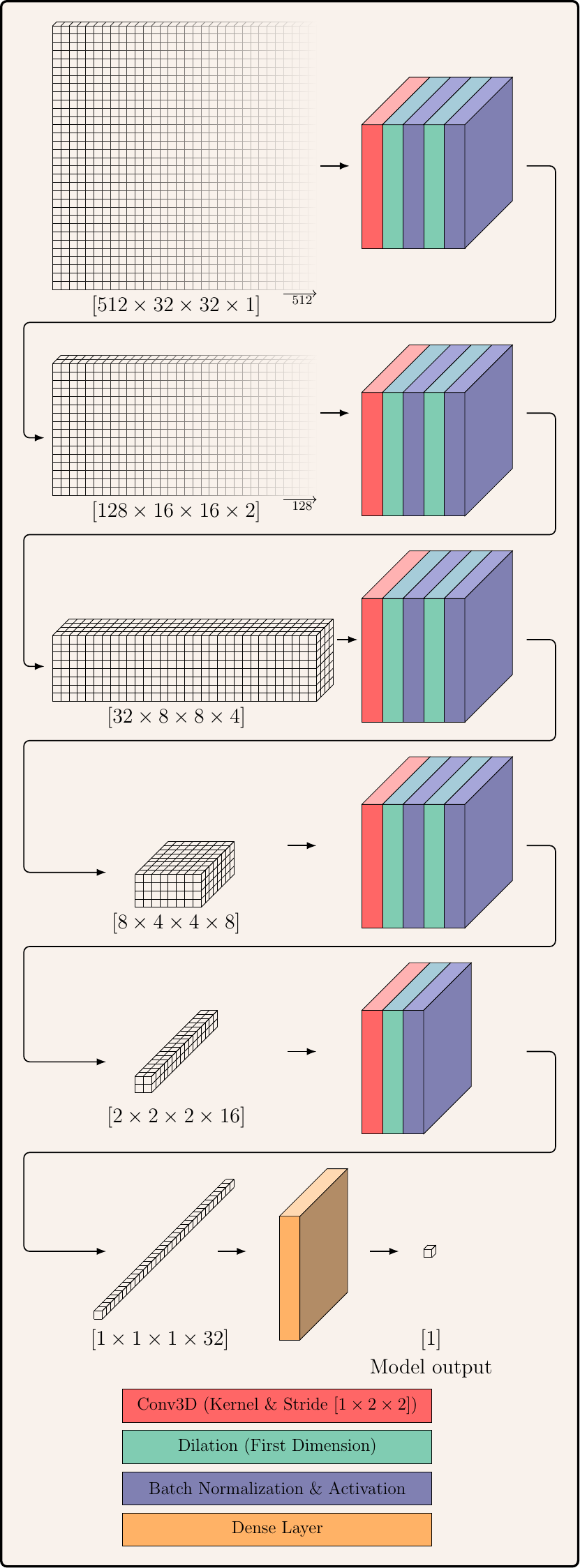}
    \caption{Architecture illustration for the model architecture. The model is used to reduce the input data from  $512 \times 32 \times 32 \times 1$ dimensions to 1. Since the second and third value for the dimensionality of that tensor are always the same, we collapse them to one dimension in the sketch. The 3D convolutional layer transforms its input dimensionality from $t \times j \times j \times k$ to $t\times \frac{j}{2} \times \frac{j}{2} \times 2k$, while the dilation is a temporally acting 3D convolutional layer with a dilation set to the powers of 2. This reduces the dimensionality in the first value by a factor of 2 each time it is called. The model output is an uncalibrated anomaly score.}
    \label{fig:CDN}
\end{figure}

After the final temporal block, 32 channels remain. A single fully connected layer maps these channels to a scalar output interpreted as an uncalibrated anomaly score.  All temporal convolutions are followed by LeakyReLU activation with $\alpha=0.1$. Figure~\ref{fig:CDN} illustrates the full architecture and intermediate dimensions. The model has 8081 parameters.

\subsection{Progressive Meta-Epoch Training Scheme}
\label{sec:training}
Training proceeds in a strictly time-forward manner to avoid information leakage  and to ensure applicability to  prospective scenarios. The timeline is partitioned into contiguous segments of fixed length $t$; for the model presented here $t = 14$ days, which was selected during the parameter search.  
Each such segment defines one \emph{meta-epoch}.  
A meta-epoch consists of: 
\begin{itemize}
    \item constructing a training set using all earlier segments,  
    \item selecting a validation set from the next chronological segment,   
    \item performing 20 ordinary epochs of gradient-based optimization, and
    \item a timeframe how often this is repeated.
\end{itemize}

Training begins once at least six meta-epochs are available, avoiding extremely small initial training sets. Within each meta-epoch, the training set includes all \EQ samples up to that time and an equal number of randomly drawn \nEQ samples. The magnitude threshold for \nEQ samples varies across parameter-search trials (Section~\ref{sec:paramsearch}).
Balancing ensures the classifier sees a broad variety of background patterns over time, while preventing the severe class imbalance inherent in seismicity.

Because older data would otherwise dominate, each meta-epoch upweights the newly added samples so that their contribution to the loss equals that of all earlier samples combined. If a segment contains no \EQ events, it is skipped as a validation window and  folded into the next meta-epoch’s training set. The validation set always contains all \EQ events in the next meta-epoch, together with a size-matched random sample of \nEQ events from that same time window.

Optimization uses the ADAM algorithm with fixed learning rate $10^{-4}$ and weight decay $10^{-5}$. A constant learning rate is required here: any decay schedule would reduce the impact of newly arriving data, and negate the whole point of the progressive training scheme. The batch-size for this model is 32, with the final batch usually being smaller.
Using this training scheme, we perform a single pass from 2001 to 2019, which we extend for the best performing models for testing to 2022.

The output of this training process is an evolving anomaly classifier whose internal parameters adapt to the changing data distribution without accessing future information.

\begin{figure} 
    \includegraphics[width=\linewidth]{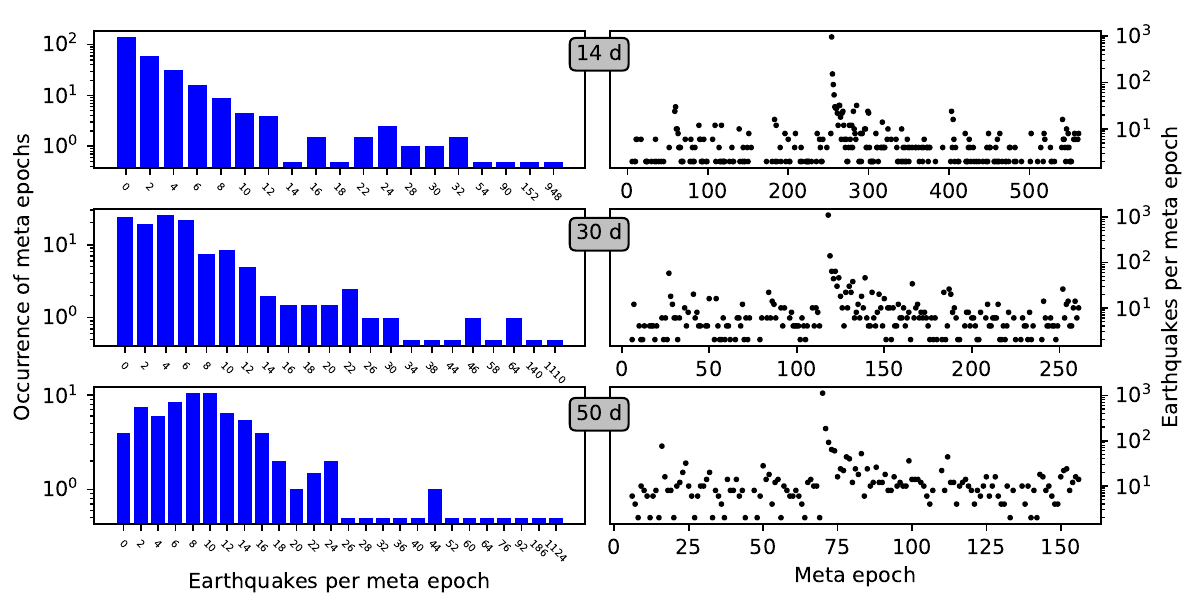}
    \caption{This figure shows the earthquake distribution in validation data. The validation is chunked in meta epochs different lengths (explained in section \ref{sec:training}).
    The left columns shows how often meta epochs with a certain number of events occur depending on the meta epoch lengths of 14, 30, or 50 days. The very high numbers correspond to the Tōhoku earthquakes and the months that follow. This can be seen in the right column, where the number of events in each meta epoch is shown over the meta epochs.}
    \label{fig:Res_Val_EQoccurrence}
\end{figure}

\subsection{Parameter Search}
\label{sec:paramsearch}

A full-factorial parameter search was conducted to assess the sensitivity of the approach to preprocessing and architectural settings.  
Each parameter combination defines a full training sweep over all meta-epochs, with final selection based on cumulative validation performance across the entire time-forward sequence. All evaluations use only the internally balanced validation sets described above. These metrics are not forecasting measures; they are used solely to select hyperparameters for the anomaly-detection model.

Table~\ref{tab:parameter_search_parameters} lists the parameters explored. The search includes preprocessing choices for \bv computation, class-definition choices for the \nEQ background, model-level choices (autoencoder on/off), and training-scheme choices (meta-epoch length, loss function). 

Computationally, each configuration requires a full temporal sweep, and training time scales approximately linearly with the number of  meta-epochs. The total number of configurations was selected to remain feasible on available hardware while still covering the interactions between preprocessing and model components.

\begin{table*}
    \centering
    \begin{tabular}{l|l|c}
         Parameter & Explanation & Possible values \\
         \hline
         dr & Radius for \bv calculation & $0.15^\circ$, $0.25^\circ$, $0.35^\circ$, $0.65^\circ$ \\
         dt & Time period to consider for \bv calculation & 150d, 365d, 730d\\
         $M_{lim}$ & Largest allowed magnitude in \nEQ & 4.0, 4.5, 4.9\\
         Loss & Loss function for the training & MAE, BCE \\
         $l_{ME}$ & Forecasting timerange / meta epoch length & 14d, 30d, 50d \\
         AE & Use of autoencoder in preprocessing  & Yes, No 
    \end{tabular}
    \caption{This table shows the parameter options for the parameter search. There are $4 \times 3 \times 3 \times 2 \times 3 \times 2 = 432$ possible combinations.}
    \label{tab:parameter_search_parameters}
\end{table*}

\subsection{Internal Validation Metrics}
\label{sec:internalmetrics}

Validation performance is computed at each meta-epoch using the balanced validation sets. We record standard classification metrics, including the area under the ROC curve (AUC), recall, precision, and the distribution of raw anomaly scores. Because the validation sets are balanced by construction, these metrics quantify model discrimination rather than prospective forecasting skill. They provide a consistent basis for comparing parameter configurations and selecting the most stable model for further use. 

For determining model performance, we look at standard classification metrics, mainly\footnote{More metrics are explained and shown in the Appendix.} the accuracy, which is defined as 
\begin{align}
    acc = \frac{TP+TN}{TP+FP+TN+FN},
\end{align}
the ratio of correctly identified time series divided by all time series looked at ($TP$ stands for true positives, $FN$ for false negatives, etc.).

\section{Results}
\label{sec:results}
This section summarizes the empirical results that guided the design of the proposed anomaly-detection method. All results shown here arise from internal validation, that is, from performance estimates obtained during the progressive time–forward training procedure described in Section~\ref{sec:training}. These diagnostics inform model architecture choices, parameter settings, and stability considerations; they do not constitute forecasting evaluation, which is beyond the scope of this paper.

\begin{figure}
    \includegraphics[width=\linewidth]{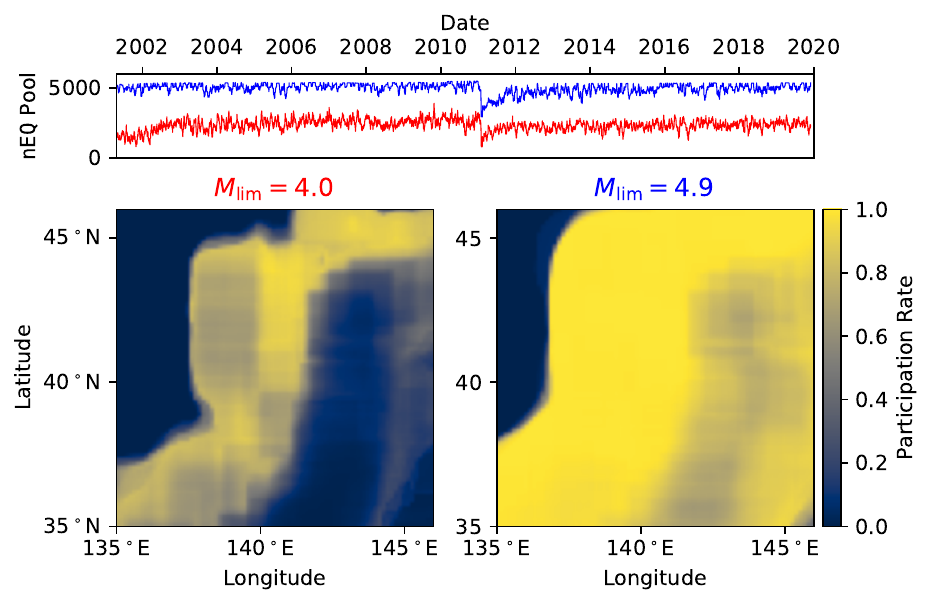}
    \caption{Availability of samples for selection as an \nEQ sample.
    The upper panel shows the availability of samples over time. For \MPerf (blue) this is noisy but constant except for the time after the Tōhoku earthquake, where a large portion of the domain is active and excluded. \MBest (red) also has less samples from the beginning of the period. The left panels shows available pool for the model with the highest training validation an overall winner of the parameter search, while the right panel shows the pool available to the best performing model with $M_\mathrm{lim} = 4.9$. Especially the region between \(35^\circ\!-\!40^\circ\,\text{N} \times 140^\circ\!-\!145^\circ\,\text{E}\) (which contains the Tōhoku earthquake) is mostly missing in the former case but readily available in the later case. An equivalent overview for all parameter configurations is Figure~\ref{fig:neq_pool_all}.}
    \label{fig:neq_pool}
\end{figure}

We call the highest-scoring model in the parameter search \MBest. It uses a magnitude limit of $M_{\mathrm{lim}} = 4$ for the exclusion criterion of the \nEQ class. This choice sharpens the contrast between positive and negative samples, which simplifies the discrimination task, but it also restricts the \nEQ pool to relatively homogeneous episodes of seismicity. As a consequence, the model learns under conditions that are easier but less representative of the full variability of real background activity. Moreover, excluding all events with $4<M<5$ leaves a physically relevant portion of the catalog unused, inviting the question of how sensitive the method is to this class-definition choice. To assess robustness against this design decision, we additionally report the best model obtained under the most permissive class definition, $M_{\mathrm{lim}} = 4.9$, which we refer to as \MPerf. This setting forces the classifier to separate large-event precursors from a much more diverse \nEQ population, which is closer to the intended use case of the method and provides a stricter test of its ability to detect meaningful spatiotemporal patterns rather than relying on artificially enhanced contrast. We therefore show results for \MPerf here and for \MBest in the appendix. The distribution of the available \nEQ pool is shown in Figure~\ref{fig:neq_pool}.

The search itself is reported in Section~\ref{sec:paramsearch_results}, but its outcomes are referenced here solely to motivate which configuration is used to illustrate training dynamics and internal validation behavior.

\subsection{Internal Validation}
During training (and concurrent validation) the data is being presented to the model in 14 day intervals, where the model is trained on some amount of data, validated on the next 14 days, and then trained on the combined data, and validated on the next 14 days. This training method is explained in more detail in Section \ref{sec:training}. Due to this specialized training procedure, giving a normal model evaluation such as the final accuracy is not as informative compared to other models. We therefore name the single training periods \textbf{meta-epochs} and report the relevant results for those meta epochs as well, instead of a single mean accuracy.

\begin{figure}
    \includegraphics[width=\linewidth]{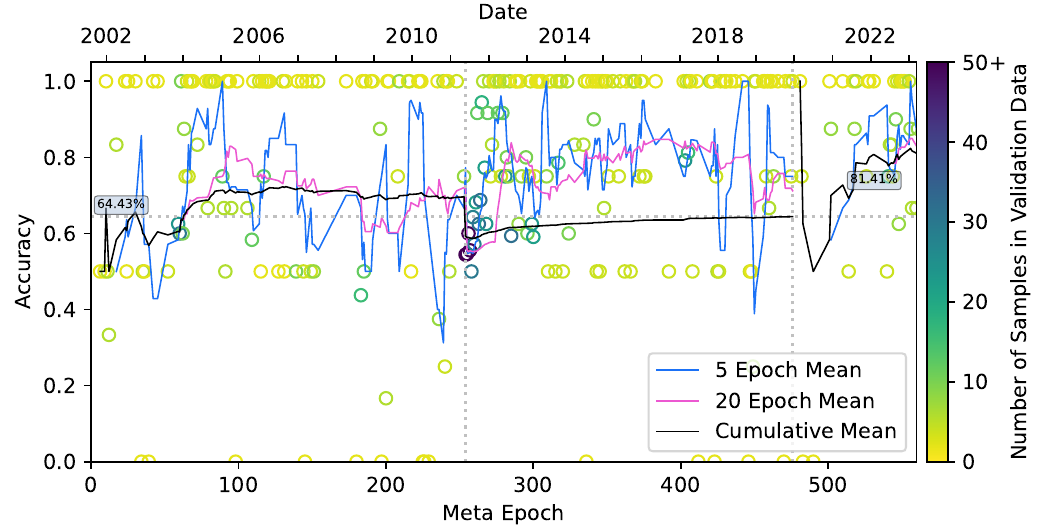}
    \caption{Classification Accuracy for meta epochs. The markers each correspond to one meta epoch and are colored based on the number of samples in the validation set for that meta epoch, however we limited the color range from 0 to 50. There are single meta epochs with much higher counts, namely those following to the Tōhoku Earthquake. A closer look at the events in the validation data is provided in Figure \ref{fig:Res_Val_EQoccurrence}. The lines show different running averages for the accuracy: The light blue line is a weighted average (since the meta epochs contain different numbers of samples),  of 5 meta epochs, the magenta line is the 20 meta epoch weighted average. Black shows the cumulative accuracy, which indicates some progression after the initial meta epochs. The two vertical dotted lines correspond to the Tōhoku earthquake and the end of the initial training period at the end of 2019. The dotted horizontal line shows the level of the training set accuracy. The solid black line after 2019 corresponds to the cumulative accuracy on the new, unseen data, where after giving a forecast for the next 14 days, the model is retrained for the next 14 days. From 2020 a new cumulative mean is shown.}
    \label{fig:Res_Acc_ME}
\end{figure}

As the progressive training changes the model and the model is applied to new data for each determination of the accuracy, the overall accuracy of the model is difficult to determine. However, accuracy is the most common metric for measuring model performance, so Figure~\ref{fig:Res_Acc_ME} shows accuracy over time. The time intervals are given in increments of 14 days, which also correspond to the progressive training cycles, the meta epochs. As the 14 day forecasting period usually contains only one or two earthquakes (see Figure~\ref{fig:Res_Val_EQoccurrence} for more details on the distribution), resulting in rather quantized and unstable accuracy values, Figure~\ref{fig:Res_Acc_ME} shows the average accuracy over 5 and 20 meta epochs as well as a cumulative accuracy.  
These running means provide a better idea of the model's performance, as the cumulative mean is dominated heavily by the varying number of earthquakes and does not recover from the Tōhoku earthquake and its aftershocks.

After selecting \MPerf based on the internal validation behavior up to 2020, we continue the progressive-training procedure through the remaining years of the catalog. The mean internal validation accuracy within the development window is $64.4\%$, reflecting how the meta-epoch scheme shapes the optimization trajectory under evolving data conditions. The portion of the curve beyond 2019 illustrates how the same training procedure behaves when applied sequentially to later data; these values are shown solely to document the stability and dynamics of the progressive-training process, rather than to assess its performance in an operational setting. The lower accuracy in the earlier meta epochs is consistent with the high concentration of events following the Tōhoku sequence, during which the model has only undergone limited training.

\begin{figure}
    \includegraphics[width=\linewidth]{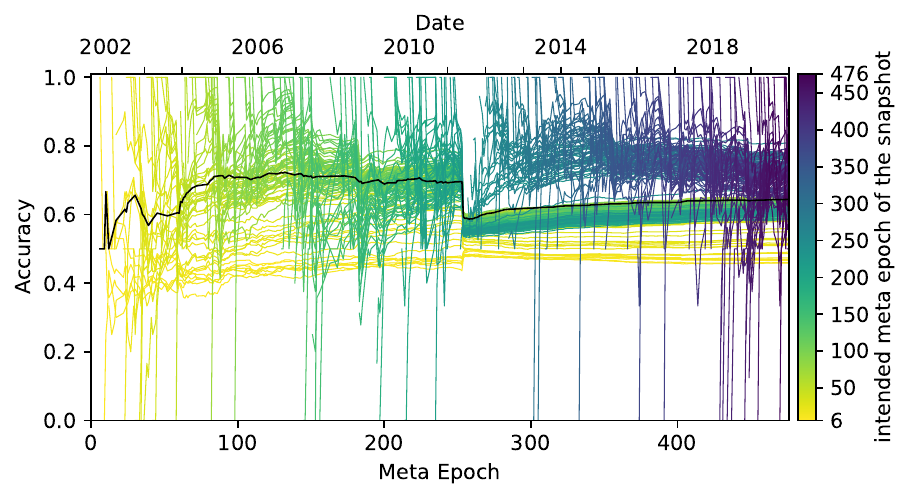}
    \caption{This figure shows the cumulative accuracy for each training snapshot after a meta epoch has finished in the colored lines, darker colors representing newer snapshots. The black line is the overall cumulative accuracy when the always latest snapshot is used.}
    \label{fig:AccuracyDecay}
\end{figure}

The part that sets this work apart from others in the field is the progressive training of the model as more data becomes available, both in this work here and in a hypothetical application where the model is continuously updated.
Figure~\ref{fig:AccuracyDecay} shows the cumulative accuracy for the model with the best performing parameter set, however, not only for the next 14 days, but rather how each snapshot of the model performs on all future data.
This clearly shows continued learning, with newer snapshots generally outperforming older snapshots. It also illustrates how the model's overall accuracy value is dragged down by the effects of the Tōhoku earthquake, as all previous models perform consistently worse there and only get a better performance very gradually (that is, they perform better long after the Tōhoku earthquake than right after. 

\begin{figure}
    \includegraphics[width=\linewidth]{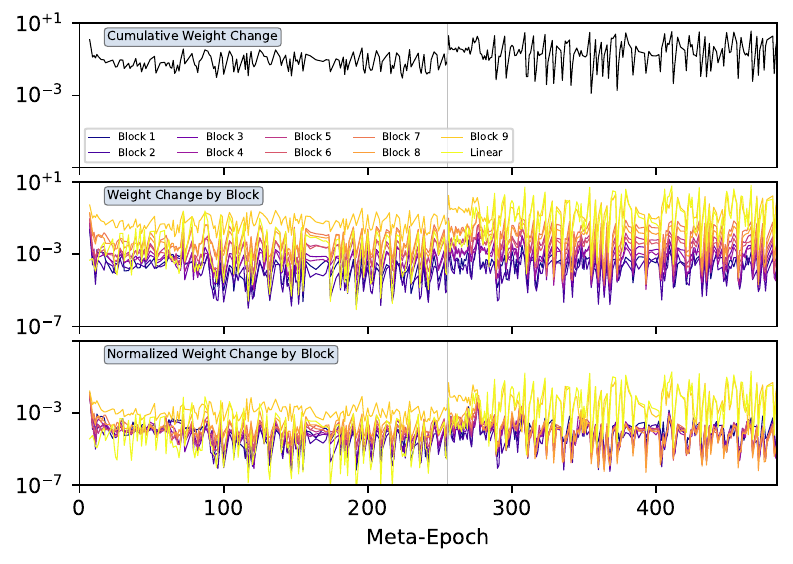}
    \caption{Weight behavior over time. The weight changes are calculated as $\sum (w_{i+1}-w_i )^2$ and displayed logarithmically. The top panel shows the overall weight change per meta epoch. The middle panel shows the weight change per block (or that of the last linear layer), and the last panels shows the weight change per block normalized by the parameter count of the block.}
    \label{fig:weight_consistency}
\end{figure}

The progressive-training scheme also allows for a detailed examination of how the network parameters evolve over time. Figure~\ref{fig:weight_consistency} summarizes the weight-change behavior across meta epochs by showing the change in parameters between meta-epochs. The change is calculated as $\sum (w_{i+1}-w_i )^2$, where for the upper panel the sum is over all parameters and for the middle panel the sum is for each block (here as a sequence of layers ending in a batch normalization, see Figure~\ref{fig:CDN}). The lower panel shows this same change per block, but it is also normalized by the number of parameters of each block. Later blocks consistently exhibit larger changes, especially after the Tōhoku earthquake, but overall patterns remain coherent across blocks, indicating that the training procedure updates the different components of the architecture in a stable and interpretable manner.

Detailed diagnostics, including ROC curves, AUC values, confusion matrices and temporal ROC variability are provided in \ref{sec:MPerf_additional}. These analyzes supplement the internal-validation picture but do not alter the selection of \MPerf or the conclusions drawn about the progressive-training procedure. The equivalent figures for \MBest are shown in \ref{sec:MBest}.

\subsection{Parameter search Results}
\label{sec:paramsearch_results}

\begin{figure}
    \includegraphics[width=\linewidth]{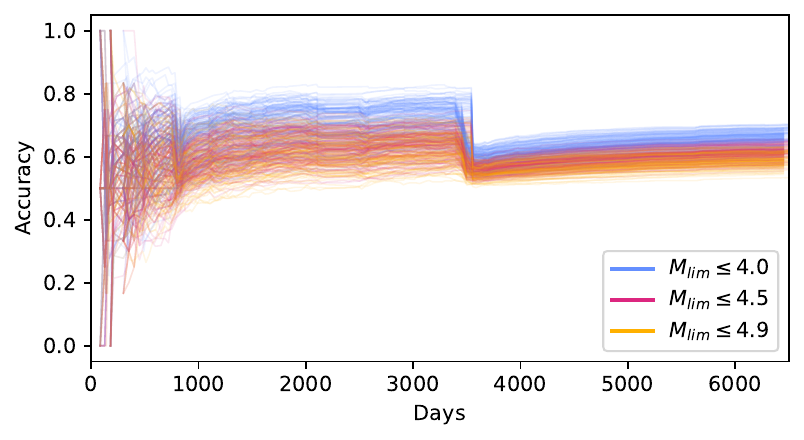}
    \caption{Validation accuracy over time. The shown accuracies refer to all models of the parameter search. As the magnitude limit for \nEQ has the biggest overall impcat, it is coloredcoded here. After the initial noise, all models exhibit a similar behavior of increase up to the Tōhoku earthquake, followed by a sharp drop and then a slow increase afterwards.}
    \label{fig:validation_accs_PS}
\end{figure}

To contextualize the choice of \MPerf, Figure~\ref{fig:validation_accs_PS} shows the meta-epoch internal-validation trajectories for all parameter combinations evaluated in the search. The dominant factor influencing variation across models is the magnitude limit used to define the \nEQ class; lower thresholds restrict the available negative samples more severely, which leads to higher internal-validation values but less stable behavior. The Tōhoku sequence introduces a marked dip for all configurations, reflecting the rapid change in data characteristics during that period. \ref{sec:app_ps} shows a detailed breakdown of the impact of all hyperparameters.

\begin{table*}
    \centering
    \begin{tabular}{c|c|c|c|c|c|c}
     & Lookback & $r$ & Maglim & $F_\mathrm{Loss}$ & ME & AE \\ 
    \hline 
    \MBest  & $730~\mathrm{d}$ & $0.15^\circ$ & 4.0 & L1 & $14~\mathrm{d}$ & OFF \\ 
    \MPerf  & $365~\mathrm{d}$ & $0.60^\circ$ & 4.9 & L1 & $14~\mathrm{d}$ & ON \\ 
    \end{tabular} 
    \caption{Results of the parameter search for the models presented in this work. ME stands for Meta Epoch and AE for Autoencoder. The results for \MPerf are presented in the main paper and the results for \MBest are in the Appendix.}
    \label{tab:PS_Models}
\end{table*}

\section{Discussion}
This work introduces a framework for detecting spatiotemporal anomalies in evolving \bvs fields using a progressive deep-learning scheme. The method combines daily \bv maps, a fixed-size spatiotemporal sampling strategy, and a forward-in-time meta-epoch training procedure. Here we discuss the behavior of the method, its technical constraints, and several data-related artifacts that influence what the model can and cannot learn from the underlying seismicity catalog. 

\subsection{Overall Method Behavior}
The progressive training design forces the model to learn representations from earlier parts of the catalog before being evaluated on later time windows. This setup avoids information leakage but also exposes the model to substantial variations in event density and \bv stability across time. A prominent example is the sharp increase in seismicity following the Tōhoku earthquake (Figure~\ref{fig:CumulativeEarthquakeNumber}). More than half of all \EQ events in the catalog occur in the multi-year aftershock sequence, and the corresponding \bv fields differ substantially from those in quieter periods.

These catalog features result in a slightly uneven distribution of positive and negative training samples across time and space (see top panel of Figure~\ref{fig:neq_pool}). During early meta-epochs, the model trains on relatively few events, whereas intermediate epochs provide disproportionately large numbers of samples from aftershock-rich regions. This imbalance shapes the internal representations learned by the model and must be kept in mind when interpreting any downstream anomalies.

\subsection{Catalog Constraints and Class Definition Artifacts}
The class labels are tied to the occurrence or absence of an \Mw $ \geq 5$ earthquake in the central region within the defined temporal horizon. This definition is simple but introduces artifacts. For example, regions with prolonged aftershock sequences generate dense clusters of positive samples, while large areas that rarely host \Mw $\geq 5$ events (while still having relevant seismic activity) contribute almost exclusively negative samples. The resulting spatial and temporal imbalance is not eliminated by balancing within meta-epochs, since the underlying distribution of \bv patterns remains strongly non-uniform.

Additionally, the \bv estimates fed into the model are computed over varying numbers of contributing earthquakes. The variability in data density, especially across tectonic regimes and between pre- and post-Tōhoku periods, leads to heterogeneity in the stability of the \bv fields themselves. These features are intrinsic to the catalog and place practical limits on the information content of the input data, but even without such extreme events, a certain heterogeneity will be present in any real earthquake catalog.

\subsection{Magnitude-Dependent Model Behavior}
\label{ssec:magnitude_dependence_discussion}
If the premise that the probability of an earthquake occurring can be forecast by using changes in \bv is correct, it would suggest that larger events are easier to forecast as they have more stress that needs to build up over a larger area.
However, the model is trained mostly on forecasting smaller earthquakes in the aftershocks of the Tōhoku earthquake, as the training data contains mostly those (see Figure \ref{fig:CumulativeEarthquakeNumber}), so it is a priori unclear how well the model works for large earthquakes.
To test this, we analyze the magnitude dependence of the network's model outputs, as shown in Figure \ref{fig:Res_Hist_all} (A). The upper plot shows the prediction distributions for different magnitude bins, the center line is marked with the actual predictions (x's), and the mean (colored) and median (black) for each bin are shown as well. 
The figure demonstrates a clear positive trend between model output and earthquake magnitude: larger magnitude events tend to produce higher model output values. This indicates that the model captures meaningful features related to event size.

\subsection{Dependence on $b$-Value Stability and Data Quality}
\label{ssec:quality_dependence_discussion}
The quality of the \bv data used for the forecasting process is bound to have an influence on the accuracy of the forecasting. Quality here is defined as the certainty with which the \bv can be determined. For this purpose, we look at the average number of earthquakes used to calculate all single values of $b$ in the input array. Note that the numbers shown are before the $b$-positive method is applied, so the actual number is roughly half.

If this value is low, the \bvs will be calculated on only a few events, which leads to drastic changes between neighboring cells in the input data, if a single earthquake is added or removed by moving the cylindrical stencil around. Theoretically, better \bvs should correspond to a better prediction; however, better or more robust \bvs will also correspond either to a higher seismicity region or a region better covered by seismic stations. However, the quality is not given to the model directly, and the model can only infer it from the way the calculated \bvs behave.

\begin{figure*}
    \includegraphics[width=\textwidth]{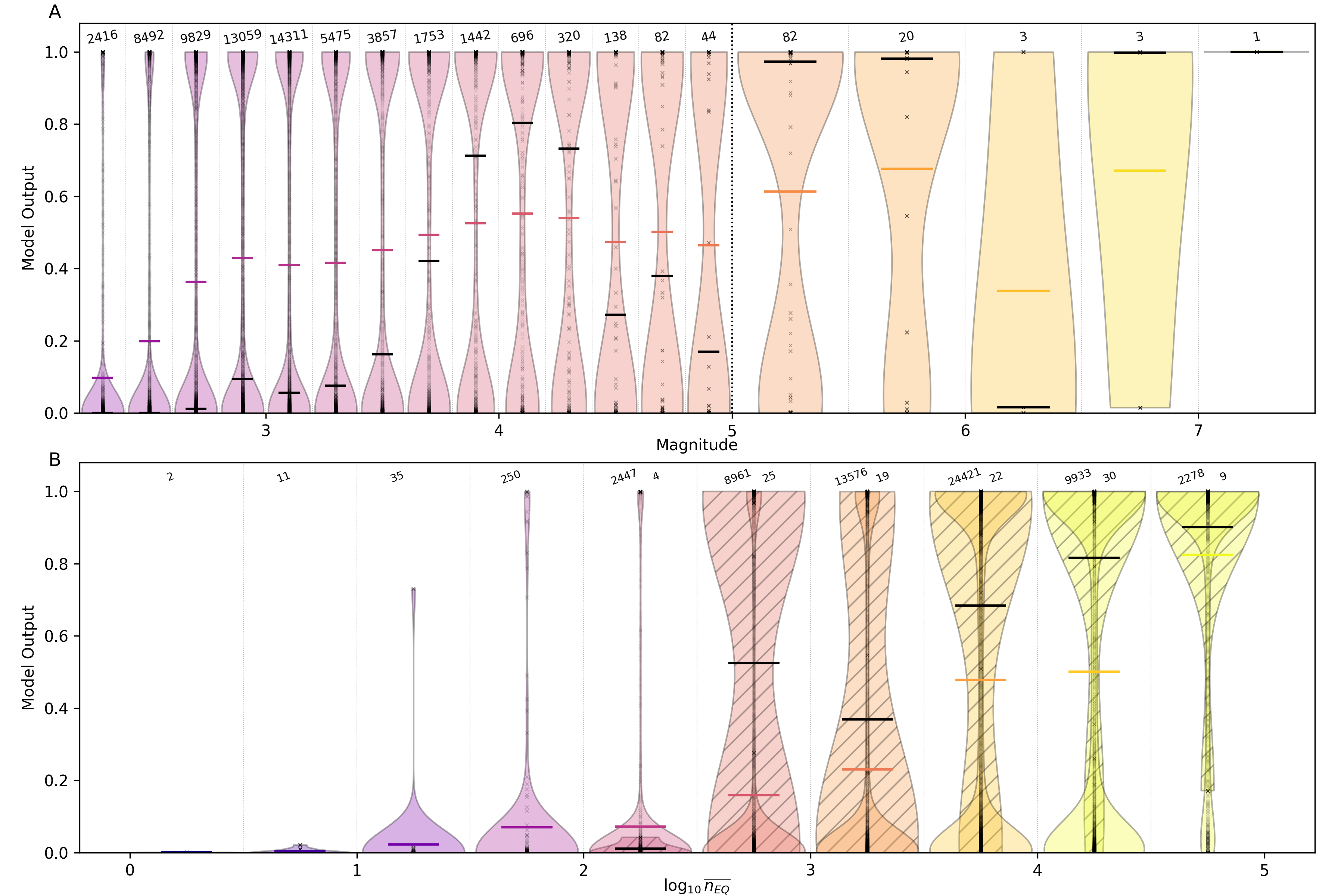}
    \caption{\MPerf output as a function of earthquake magnitude and seismicity level. The upper panel shows average model output (horizontal bars) and violin plots of model output with respect to event magnitude for the test set (post-2019). The number of events per magnitude bin is indicated above each violin. A clear increase in model output with event magnitude is observed. The lower panel displays model output as a function of the average number of earthquakes contributing to \bv calculation. Hatched violins and black vertical lines represent time series ending with events of magnitude \Mw $\geq$ 5 (\EQ), while colored violins and colored horizontal lines represent \nEQ time series (\Mw $<$ 5). Sample counts for both classes are shown above each bin (left: \nEQ, right: \EQ).}
    \label{fig:Res_Hist_all}
\end{figure*}

The impact of this value is shown in the lower panel of Figure \ref{fig:Res_Hist_all} (B). It shows the model output as a function of the average number of earthquakes used in the corresponding \bv time series calculation. The model distinguishes between the \EQ and \nEQ classes even for similar levels of seismic activity. This is evident from the consistent separation of the mean outputs in each bin, with \EQ samples (hatched, area, black vertical lines) systematically exhibiting higher model output than \nEQ samples (clear area, colored vertical lines).

\subsection{Limitations and Methodological Implications}
Several features of the method impose constraints that future work may need to address. The fixed-size 512-day temporal window forces the model to represent long-term \bv evolution, but also mixes multiple stages of seismic sequences into each sample. Furthermore, the progressive training procedure exposes the model to temporal concept drift: the distribution of \bv patterns after the Tōhoku earthquake differs markedly from the distribution before it, and the aftershock sequence is also continuously ebbing. 

More broadly, the reliance on \bv alone means that the method can only detect anomalies expressible through changes in Gutenberg–Richter statistics. Other precursory signals such as geodetic strain, stress changes, or cascading triggering effects are not represented directly. Thus the anomaly scores should be interpreted as indicators of patterns in \bv evolution, not as comprehensive assessments of evolving seismic hazard.

\subsection{Summary}
The analysis presented here characterize how the anomaly-detection model behaves under real catalog conditions. The magnitude- and quality-dependent effects highlight the extent to which the model's outputs reflect not only genuine spatiotemporal patterns but also properties of the catalog and \bv estimation process. These structural issues represent methodological limitations that must be considered when applying the model in any downstream context.

\section{Conclusions}
In this work we presented a progressive training framework that uses spatiotemporal \bv distributions to derive anomaly scores associated with impending large earthquakes. The method operates on a limited catalog and learns from gradually expanding training windows, enabling the model to adapt to changes in regional seismicity while preserving chronological integrity. This design makes it possible to study how model behavior evolves as additional data become available, and how catalog features—such as completeness, event density, and large-sequence dominance—shape the learned representations.

The analyses demonstrate that the model's outputs are strongly conditioned by properties of the input catalog rather than purely by physical precursors. Two effects are particularly notable. First, the dominance of the 2011 Tōhoku earthquake and its extensive aftershock sequence introduces a substantial imbalance in the training data, influencing the distribution of anomaly scores across the entire temporal window. Second, both event magnitude and the number of earthquakes contributing to \bv estimates introduce systematic structure in the model outputs: larger events and more stable \bv estimates tend to correlate with higher anomaly scores. These patterns are consistent with artifacts that arise from data availability and spatial–temporal sampling, and they highlight the importance of understanding input-driven effects before interpreting any model behavior in physical terms.

Taken together, the results show that deep learning models operating on representations based on \bv must be interpreted with caution. The method presented here identifies patterns in the \bv field that correlate with large events, but those patterns are inseparable from underlying catalog characteristics, including completeness variations, nonuniform seismicity, and the presence of dominant sequences. The approach therefore provides a structured way to explore such relationships and to quantify how methodological choices influence anomaly detection, rather than serving as a calibrated forecasting tool in its current form.

\section{Future Work}
This method has three general directions in which it can and should be developed.
First, an adaption to more rate-forecast like output can be done in a number of ways, including rescaling or adjusting the target during training to represent a rate rather than an anomaly score. 

Secondly, there are more methodological adjustments that can be made, but as the choice there is quite large, they were not included here. Possible adjustments include a ``forgetting'' of older events by removing them from the training data, a time-since-event dependent weight for samples, inclusion of more input channels or a different treatment of the class imbalance.

Thirdly, the method could be tried on different seismic regimes, California or Italy come to mind there.

\section{Acknowledgments}
This research is supported by the ``KI-Nachwuchswissenschaftlerinnen'' -- grant \texttt{SAI 01IS20059} by the Bundesministerium für Bildung und Forschung -- BMBF. The calculations were performed at the Frankfurt Institute for Advanced Studies'  GPU cluster, funded by BMBF for the project Seismologie und Artifizielle Intelligenz (SAI). We thank Megha Chakraborty, Darius Fenner, Dr.~Claudia Quinteros, Professor Geoffrey Fox, Dr.~Danijel Schorlemmer and Dr.~Kiran Thingbaijam for their helpful discussion. 
We acknowledge the help and advice from Prof.~Dr.~Horst Stoecker.
The research has made extensive use of PyTorch \citep{pyTorch2019}, numpy \citep{Numpy2020}, and matplotlib \citep{Matplotlib2007}.
Geographical maps were made with Natural Earth. Free vector and raster map data @ naturalearthdata.com.
The training was carried out on Nvidia A100 Tensor Core GPUs.

\section{Open Research}
The codes used in this study for pre-processing and classification are available at https://zenodo.org/doi/10.5281/zenodo.10829447

\bibliography{bibliography.bib}
\newpage
\appendix
\setcounter{figure}{0}
\setcounter{table}{0}
\section{Autoencoder Approach}
\label{sec:AE}
The idea from \citep{Fox2021} on which the use of the autoencoder here is based is, that the autoencoder learns the ``normal'' state of the system, so when the system is in an abnormal state, the reconstruction error will be different. A second network, the main classifier, will then be used on the reconstruction error to classify the input between ``normal'' (\nEQ) and ``abnormal'' (\EQ) states.
Empirically we have found, that using this two stage architecture sometimes works better than using the classifier immediately on the raw data, the exact extent of this will be shown in the section on the parameter search.

\begin{figure}
    \includegraphics[height=0.85\textheight]{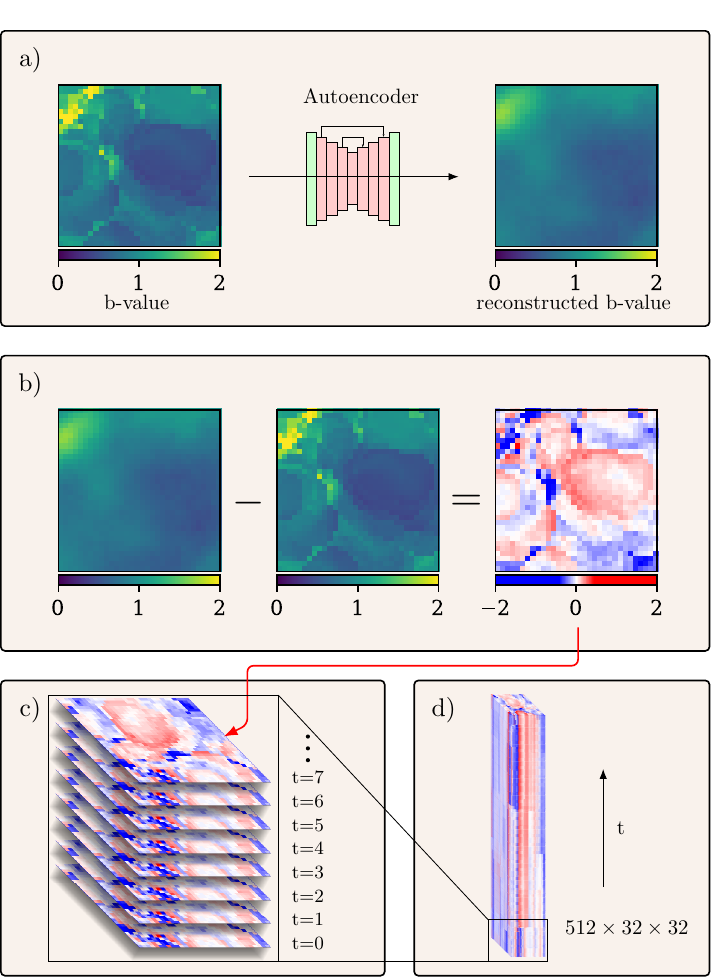}
    \caption{Pipeline from the \bv grid to the input of the classifier network when used with the autoencoder. First (a) all relevant $32 \times 32$ maps of the \bv are sent through the autoencoder. In the second step (b) we take the difference between reconstruction and input to get a difference map. In the third step (c), 512 different maps are assembled in a sequential manner to form the final input block shown in (d).}
    \label{fig:BlockCreation}
\end{figure}

If the autoencoder is used, we process the $[32 \times 32]$ \bv array and run it through an autoencoder as a first step.
We then take the difference between reconstruction and input to create an array of reconstruction error. This is done for all 512 instances per block to create the input for our second step. The process is also illustrated in Figure~\ref{fig:BlockCreation}.

The autoencoder is trained on a dataset of 150,000 images of $32 \times 32$ pixel values of the \bv from the region of which samples could be taken in for the classifier following the \nEQ criterion. This includes all lookback + 512 days before an earthquake at that location (which would then be centered in the $32 \times 32$ image) as well as all locations that can possibly be selected according to the \nEQ criterion defined above. $80\%$ of those images are used as training data, $20\%$ are used for testing.

The training is performed using mean squared error as the loss, and the model is trained until the validation loss does not improve for 50 epochs. 
We tried out different architectures for this task, and a network based solely on 2D convolution layers performed best.

\begin{figure}
    \tikzstyle{block} = [draw, fill=red!20, rectangle, minimum height=2em, minimum width=8em]
    \tikzstyle{coord}  = [coordinate, on grid]
     \begin{tikzpicture}[node distance=2em,>=latex', scale=0.4]
        \node [block, minimum width=18em, fill=green!20](d0) {Input: $32 \times 32$, $1$ channel};
        \node [block, minimum width=16em, below of=d0]  (d1) { $4$ channels};
        \node [block, minimum width=14em, below of=d1]  (d2) { $8$ channels};
        \node [block, minimum width=12em, below of=d2]  (d3) { $16$ channels};
        \node [block, minimum width=10em, below of=d3]  (d4) { $32$ channels};
        \node [block, minimum width=12em, below of=d4]  (d5) { $16$ channels};
        \node [block, minimum width=14em, below of=d5]  (d6) { $8$ channels};
        \node [block, minimum width=16em, below of=d6]  (d7) { $4$ channels};
        \node [block, minimum width=18em, below of=d7, fill=green!20]  (d8) {Output: $32 \times 32$, $1$ channel};
        \node[coord, right= 12em of d1] (h1a) {};
        \node[coord, right= 10em of d3] (h2a) {};
        \node[coord, right= 10em of d5] (h2b) {};
        \node[coord, right= 12em of d7] (h1b) {};
        \draw [->] (d1) -- (h1a) -- (h1b) -- (d7);
        \draw [->] (d3) -- (h2a) -- (h2b) -- (d5);
    \end{tikzpicture}
    \caption{This figure shows the architecture of the autoencoder for the \bv arrays, with skipped connections.}
    \label{fig:AE_Architecture}
\end{figure}
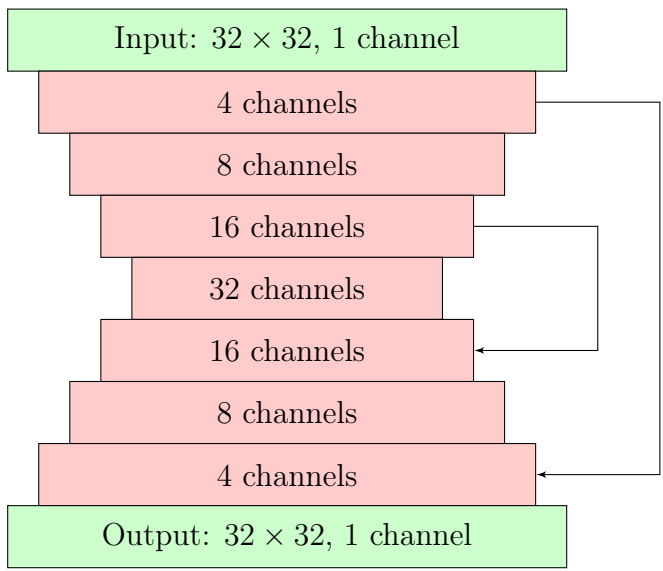

The final model consists of 8 layers, 4 for the encoder part and 4 for the decoder, where the number of channels doubles each layer (encoder) and is cut in half (decoder), except for the first layer in the encoder, where the output consists of 4 channels, and the last layer of the decoder, where the channels are reduced from 4 back to 1. There are skipped connections, as shown in Figure~\ref{fig:AE_Architecture}.

\newpage
\section{Additional Data Diagnostic Figures}

\begin{figure}
    \includegraphics[width=\linewidth]{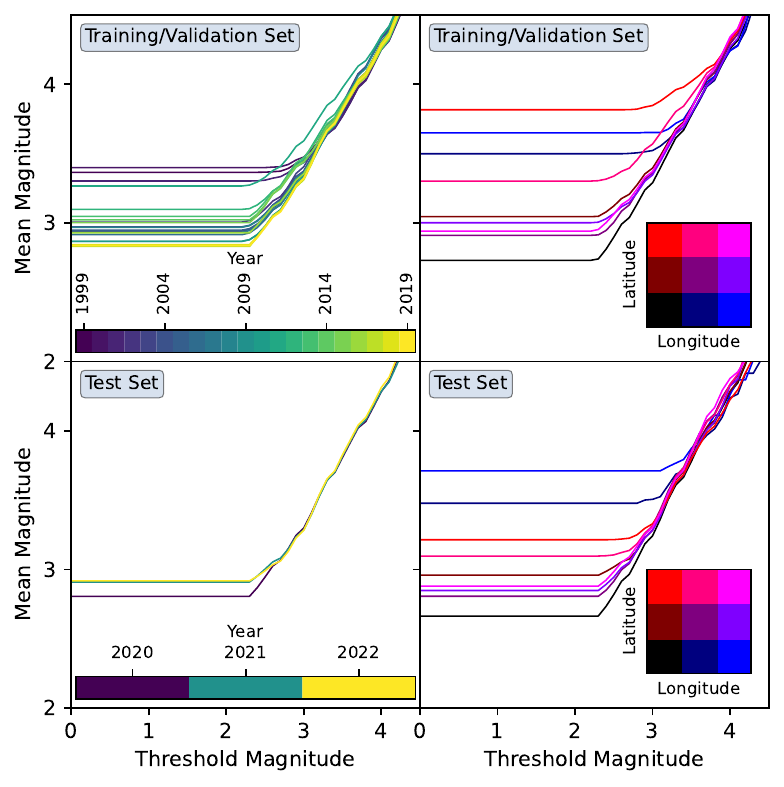}
    \caption{Magnitude Completeness using the method presented in \citep{Godano2023} for training and validation (top) and the test set. Left panels show an analysis by year. This shows a general trend of improvement, with 2011, the year of the Tōhoku earthqauake standing out again. The right panels show a completeness analysis based on location, where we divide our $[35^\circ - 46^\circ \mathrm{N}] \times [135^\circ - 146^\circ \mathrm{E}]$ region into 9 equal sized squares and show the completeness for all of them, space encoded in the color square as shown in the figure.}
    \label{fig:completeness}
\end{figure}

\begin{figure}
    \includegraphics[width=\textwidth]{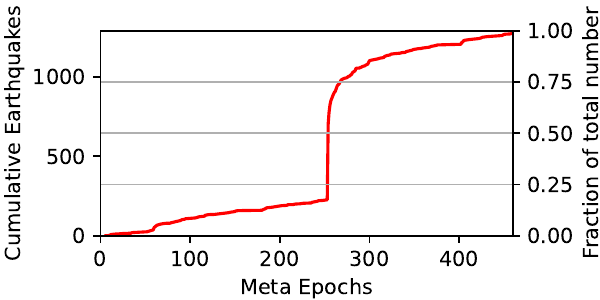}
    \caption{This figure shows the cumulative number of earthquakes in the training set. The meta epoch length is 14 days. Roughly extrapolating the line before the Tōhoku earthquake, we can estimate that around 60\% of the earthquakes in the dataset are aftershocks, skewing the training data strongly towards those.}
    \label{fig:CumulativeEarthquakeNumber}
\end{figure}

\begin{figure*}
    \includegraphics[width=\linewidth]{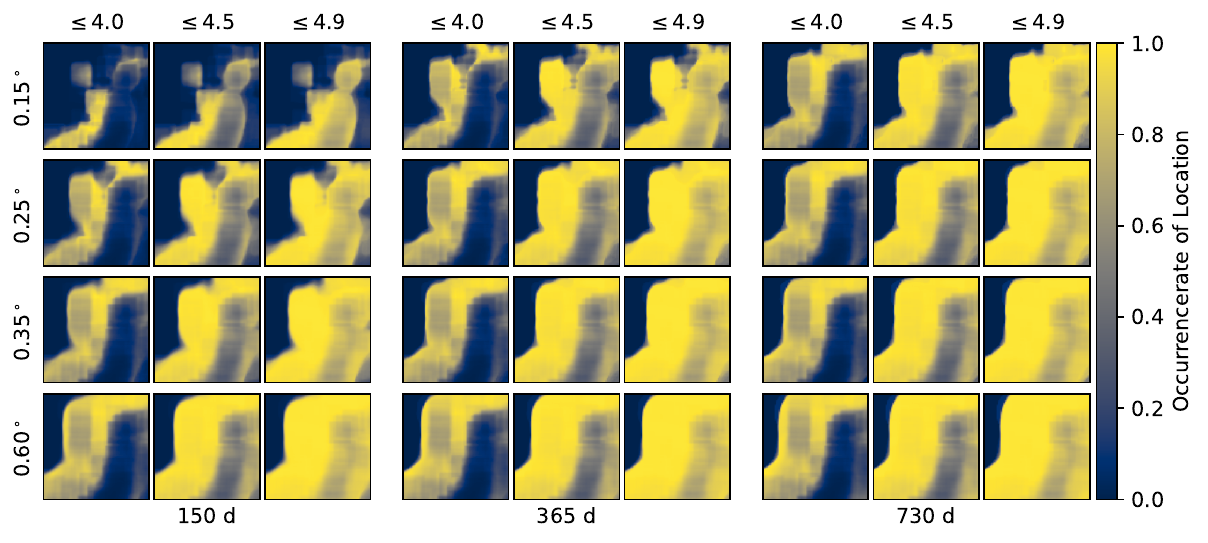}
    \caption{Availability of samples for selection as an \nEQ sample for all parameter configurations that affect \nEQ. \MBest takes the top left map of the right block, \MPerf the bottom right map of the central block.}
    \label{fig:neq_pool_all}
\end{figure*}

\newpage
\section{Parameter search details}
\label{sec:app_ps}
In order to determine many of the seismologically sensitive parameters in our model, we performed a parameter search to determine the best performing models. The parameters we considered are shown in Table\ref{tab:parameter_search_parameters}.

The first two parameters are the radius for \bv calculation and the time period for the \bv calculation; as both influence the \bv, we will have $4 \times 3 = 12$ different \bv datasets.
The third parameter determines the magnitude limit for exclusion of events from the \nEQ class, so that in our classification we do not punish the model for misclassifying events that are close to our threshold of 5; the reason is twofold: On the one hand, the magnitude conversion introduces an uncertainty, so some events might be mistakenly ignored or included in our \EQ class. On the other hand, this might increase the ``contrast'' between the two classes, which improves performance, while also making the model less applicable to operational application.

These three first parameters make for 36 parameter configurations which, to be compared fairly, each require their own autoencoder, as the definitions of \nEQ influence the data on which the autoencoder is trained. All of these autoencoders have the same architecture.
Their training data is chosen randomly from all possible locations where the \nEQ condition is met (See Figure~\ref{fig:neq_pool_all}). Since the number of positions meeting this condition varies widely based on the magnitude limit, we chose 150000 training samples, which is a number all 36 cases can meet.

The last three parameters only affect the classifier.
The fourth parameter sets the loss function to be either the Mean Average Error (MAE) or Binary Cross Entropy (BCE). As our model only has one output per sample, MAE is just the linear error.

The fifth parameter is the meta-epoch length time frame. We therefore have different total meta epochs, 484, 255, and 134 (until the end of 2019). 

Lastly, we test the method proposed in \citep{Fox2022} by using an autoencoder on our data first, so we run all parameter configurations once with the autoencoder and once without.

\subsection{Parameter search results}
\label{ssec:parametersearch}
\begin{figure}
    \includegraphics[width=\linewidth]{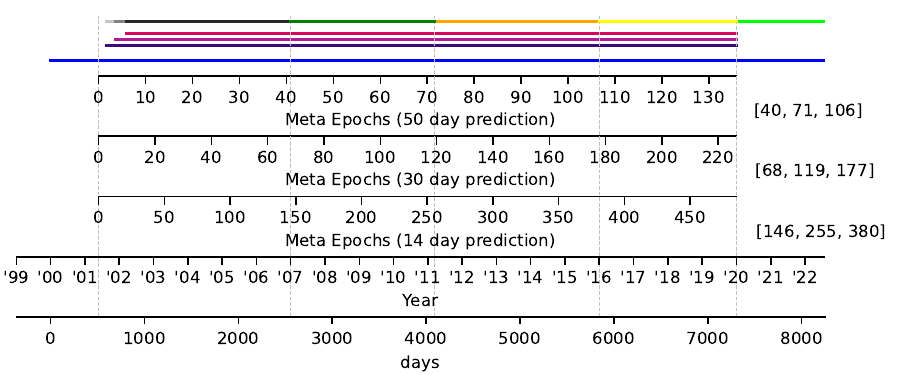}
    \caption{This figure shows the different units and instances of important times for this work and more specifically for the parameter search. The \bv is calculated for each day marked by the blue line. Consequently, the block which we use as our input of 512 \bv maps can start 512 days later. The 6 meta epochs after which we start training are shown with the differing starting location of the three purpelish lines, each corresponding to one of the three meta epoch lengths (From top to bottom: 50d, 30d, 14d). The segmented line at the top shows the different periods, each color representing one of them. For our final evaluation we can use the light green period, which has not been used in the parameter search. Lower down the correspondence of the meta epochs to dates is shown, and on the right next to each of those lines we show the three meta epochs which contain the cutoff date between the different periods.}
    \label{fig:PS_Timeline}
\end{figure}

For the parameter search we present three figures of increasing detail.
In all cases, we divided up the catalog in four time segments:
From start (which differs due to the different lengths of meta epochs) to the end of 2006 it period 1. During this time, the model just training, and while there is some output, this should be taken with a grain of salt.  From 2007-01-01 to 2011-03-10 (the date of the Tōhoku earthquake) is period 2, during which we can assume that the model is trained to some degree. From 2011-03-11 to 2015-12-31 is period 3, which is dominated by the Tōhoku earthquake and its aftershocks, which are visible in the data for many years. Period 4 is from 2016 onward, which contains less activity spawned by the Tōhoku earthquake (though not none). The goal with this fourth period is twofold: For one it helps to separate out the dominance of the Tōhoku event from metrics based on number of earthquakes, since most periods containing the months after the Tōhoku earthquake will be numerically dominated by them. Two, Tōhoku earthquake and the enduring aftershocks form a better basis for training, as they account for roughly $60\%$ of all data in the training (refer also Figure \ref{fig:CumulativeEarthquakeNumber}).
A visual guide for the segments is shown in Figure~\ref{fig:PS_Timeline}, which also includes the meta epochs for the three different meta epoch lengths.

\begin{figure}
    \includegraphics[width=\linewidth]{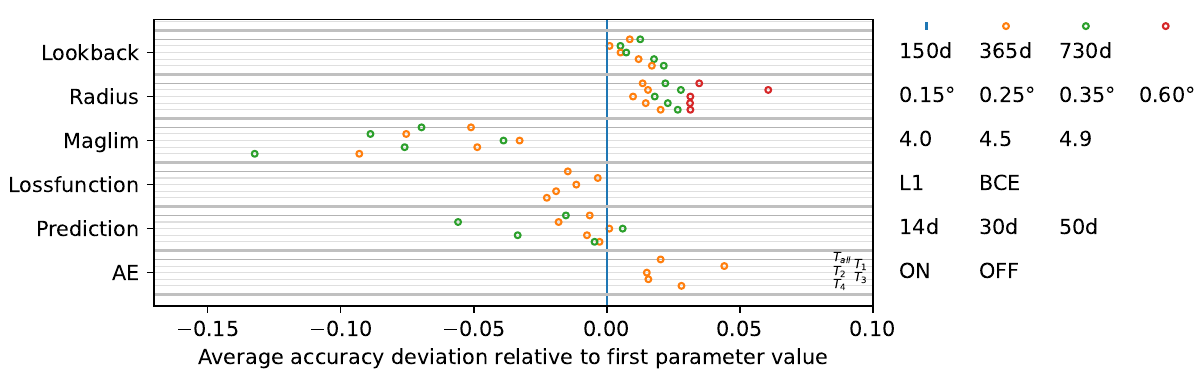}
    \caption{This figure shows the impact of varying the different parameters on the accuracy relative to one another. The first entry of the table on the right is taken as the baseline (blue) and the deviation from that baseline are shown as circles in the other colors corresponding to the columns the parameter in question is in. The different horizontal lines within any one parameter-block correspond to (from top to bottom) the result for the whole time domain and then the consecutive time segments below that.}
    \label{fig:PS_res_1of3}
\end{figure}

First, Figure~\ref{fig:PS_res_1of3} shows the average deviation in accuracy when varying only that parameter. This shows the sensitivity to certain parameters, namely a high sensitivity to the magnitude limit for the \nEQ class, while the loss function only has a minor impact.
The optimal parameter configuration based off this figure is $[730\mathrm{d}, 0.6^\circ, 4.0, \mathrm{MAE}, 14\mathrm{d}, OFF]$.

\begin{figure*}
    \includegraphics[width=\textwidth]{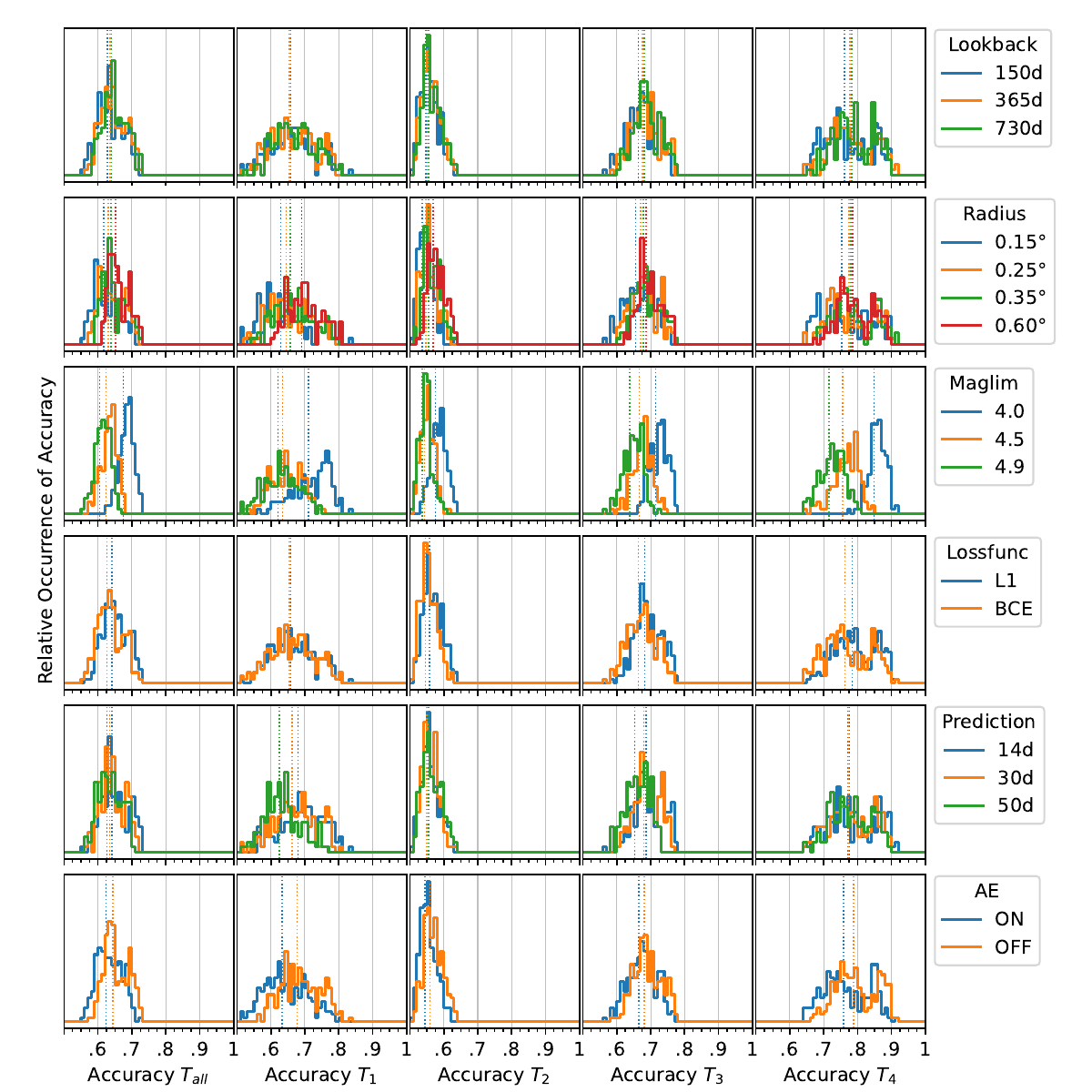}
    \caption{This figure shows histograms for the different parameters the parameter search and for the different time periods of the training interval. The rows correspond to the parameters on the right, the columns (from left to right) correspond to the time periods. The first column shows the accuracies over the whole time period, after that it is periods 1-4 as shown in Figure~\ref{fig:PS_Timeline}.}
    \label{fig:PS_res_2of3}
\end{figure*}

Figure~\ref{fig:PS_res_1of3} can be made more detailed by providing a histogram for the accuracy distribution of each parameter, which is shown in Figure~\ref{fig:PS_res_2of3}. In addition to the accuracy distribution, the vertical lines show the mean for each parameter. The rows correspond to the parameter being varies (indicated on the right), and the columns correspond to the time segments.

\begin{figure}
    \includegraphics[height=0.9\textheight]{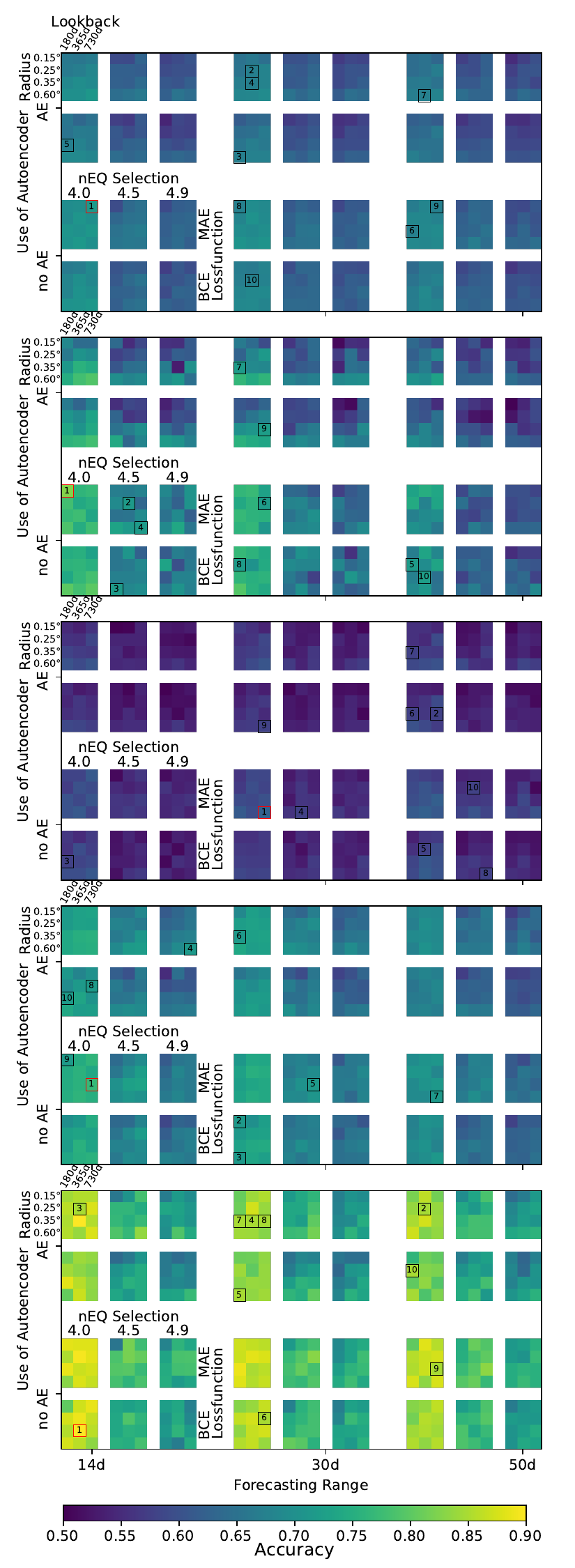}
    \caption{ This figure shows the accuracy for all runs of the parameter search, where the top ten accuracies are marked with $\square$. 
    The figure can be read as follows (using the upper panel as an example): The best overall model is in the top panel, marked with a red $\square$. It is in the lower two rows, so no Autoencoder was used, and within those lower two rows it is in the upper one, confirming that indeed, MAE was the loss function. It is in the first block of columns, so the forecasting range was 14 days, while in its block of columns it is in the first column, corresponding to a magnitude limit of 4.0 for the \nEQ selection. Now, in the small $4 \times 3$ block, it is in the first row (radius for \bv calculation  $0.15^\circ$) and the third column (lookback = 730 d).
}
    \label{fig:PS_res_3of3}
\end{figure}

Lastly, Figure~\ref{fig:PS_res_3of3} shows the accuracies for all parameter configurations and all time segments. The time segments are represented by the different panels, starting from the top with $T_{all}$ and $T_1$, $T_2$, $T_3$, and $T_4$ below it in that order. The results here broadly correspond to what can be seen in the two previous figures, i.e.\ MAE outperforms BCE and a lower magnitude threshold for the classification of \nEQ improves the classification. The best overall result is $[730\mathrm{d}, 0.15^\circ, 4.0, \mathrm{MAE}, 14\mathrm{d}, OFF]$, differing in \bv calculation radius from what Figure~\ref{fig:PS_res_1of3} suggested. 

\subsection{Discussion of the parameter search results}
The result of the parameter search when looking at each model individually yields a slightly different result than what is expected from the mean results. As for the whole project, the number of events on which the accuracy is based is quite low, so there is a good chance that, in principle, over a long time with more data, another model performs better. However, the accuracy is quite similar for all the best models, and this work is not meant for immediate implementation of a Deep Learning algorithm for earthquake forecasting, rather as a proof of concept for the overall method. The result in general and the tendency in which the parameter search points should be taken more seriously than the efficacy of the resulting best model. 

\section{Additional Performance Metrics}
\label{sec:MPerf_additional}
\begin{figure}
    \includegraphics[width=\textwidth]{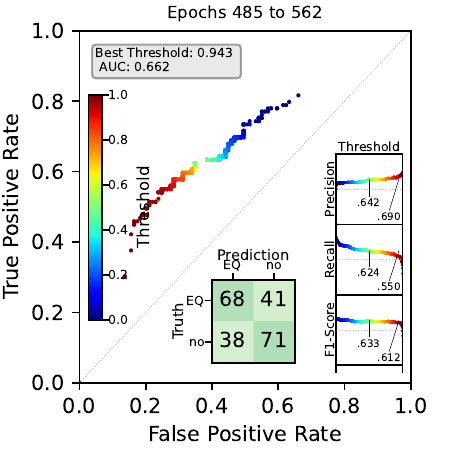}
    \caption{Receiver Operating Characteristic (ROC) curve and classification metrics for the best-performing model, evaluated on a balanced dataset from 2020 to 2022. The optimal threshold derived from the ROC curve is 0.837; however, as shown in the inset plots (axes scaled 0–1), performance at the fixed threshold of 0.5 remains close to optimal. The confusion matrix summarizes classification results at the 0.5 threshold.}
    \label{fig:ROC_etc}
\end{figure}

As the accuracy can give a skewed picture of the performance, there are other metrics that help diagnosing model performance:
\begin{align}
    \mathrm{Precision} &= \frac{TP}{TP+FP}\\
    \mathrm{Recall} = TPR &= \frac{TP}{TP+FN}\\
    FPR &= \frac{FP}{FP+TN}\\
    F1 &= \frac{2}{\mathrm{Precision}^{-1} + \mathrm{Recall}^{-1}} = \frac{2 TP}{2TP + FP + FN}
\end{align}
where $TPR$ and $FPR$ stand for True Positive Rate and False Positive Rate.
Those two quantities are also used in the Receiver Operator Characteristic (ROC), where $TPR$ is shown against $FPR$ while sliding a classification threshold across the outputs, as seen in Figure \ref{fig:ROC_etc}. Therefore, the tradeoff between $TPR$ and $FPR$ can be evaluated in relation to the threshold and determined in relation to the problem at hand. The overall performance can be estimated using the Area Under the Curve (AUC) of the graph, where a higher AUC corresponds to a better model.

The ROC curve is commonly used to determine an optimal classification threshold by evaluating the trade-off between true-positive and false-positive rates.
However, in our case, identifying a single optimal threshold from the full evaluation domain is not straightforward. The training data includes periods with varying seismic activity, notably the early phases of training and the prolonged aftershock sequence following the Tōhoku earthquake. These temporal variations lead to substantial shifts in the estimated optimal threshold when ROC curves are calculated for different time intervals: while the model consistently achieves an AUC above 0.5, optimal thresholds range from 0.02 to 0.99, depending on the period analyzed (Figure~\ref{fig:ROC_epochs}).
Given this variability, we adopted a fixed classification threshold of 0.5, which provides robust and stable performance across different eras. This choice avoids overfitting to particular temporal features of the data while maintaining acceptable trade-offs in precision, recall, and F1-score, as shown in Figure \ref{fig:ROC_etc}. We emphasize that although ROC analysis is typically not performed directly on test data (as choosing a threshold on the test data distorts the results), we include this evaluation to demonstrate that our threshold selection approach does not materially degrade performance.
For this analysis, we used the time series preceding the \Mw $\geq 5$ earthquakes from 2020 to 2022 (see Figure~\ref{fig:PS_Timeline}, lime green interval), along with an equal number of \nEQ time series selected with a magnitude threshold of 4.9.

\subsection{ROC Analysis Over Time}
\begin{figure*}
\includegraphics[height=0.9\textheight]{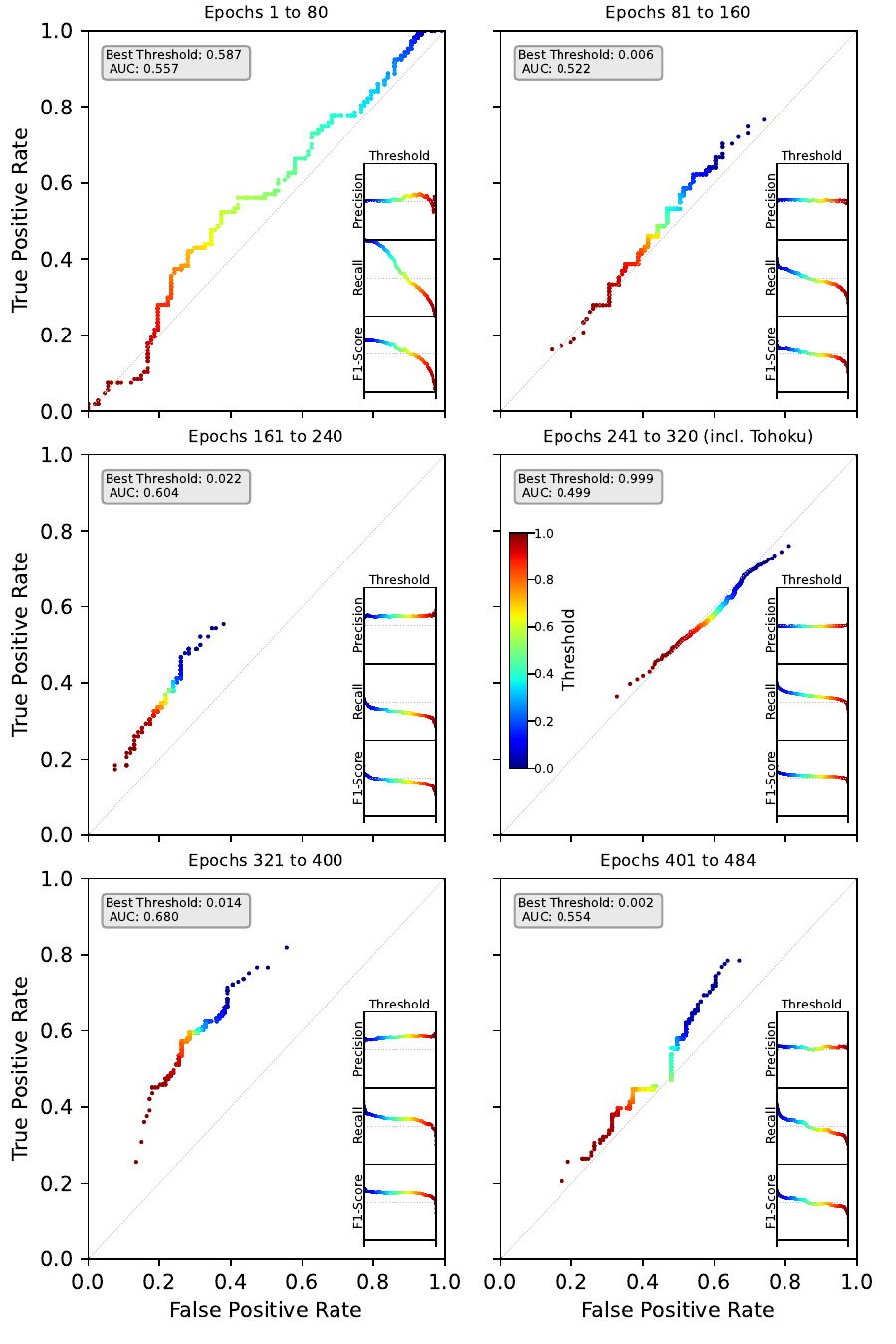}
\caption{Receiver Operating Characteristic (ROC) curves for different training intervals. The training period is divided into six approximately equal-length segments, chosen to match the duration of the interval used in Figure \ref{fig:ROC_etc}. A marked change is observed between intervals preceding and following the 2011 Tōhoku earthquake (occurring at meta epoch 255, near the beginning of the center-right panel). The influence of this major event is evident in both the panel containing the Tōhoku earthquake and the subsequent two panels, where substantially lower optimal thresholds are observed compared to the earlier intervals. Despite these shifts, a fixed threshold of 0.5 remains robust across all intervals.}
\label{fig:ROC_epochs}
\end{figure*}

We chose not to adjust the classification threshold dynamically based on ROC analysis. As illustrated in Figure~\ref{fig:ROC_epochs}, the optimal threshold varies significantly depending on the time interval analyzed. However, a fixed threshold of 0.5 consistently provides performance that is not substantially worse than the optimal value for any given period, while a given optimal value for one period might perform worse on another. This temporal variability in the optimal threshold is one of the reasons for selecting a constant cutoff value.

\newpage
\section{Model performance for \MBest}
\label{sec:MBest}

\label{sec:MBest}
In the main analysis, we follow \MPerf, the model with the highest accuracy during training, while also having a magnitude limit of 4.9. This magnitude limit decreases performance during training, as there are more hard-to-decide cases in the training data, a feature that is desirable for realistic applications. The model that had the best training performance, \MBest performs worse under realistic conditions, as we will show here.

\MPerf and \MBest were chosen solely on their performance during training (in case of \MPerf by including the additional condition of maglim = 4.9 in the parameter search). We did not test all models of the parameter search to find the one that performs best on a realistic test set, as this would unacceptably leak information from the test set into model selection.

The following figures all have analogues to previous figures in this work. Each figure here contains a reference to that previous figure as well as a brief comparison of the contents.

\begin{figure*}
    \includegraphics[width=\textwidth]{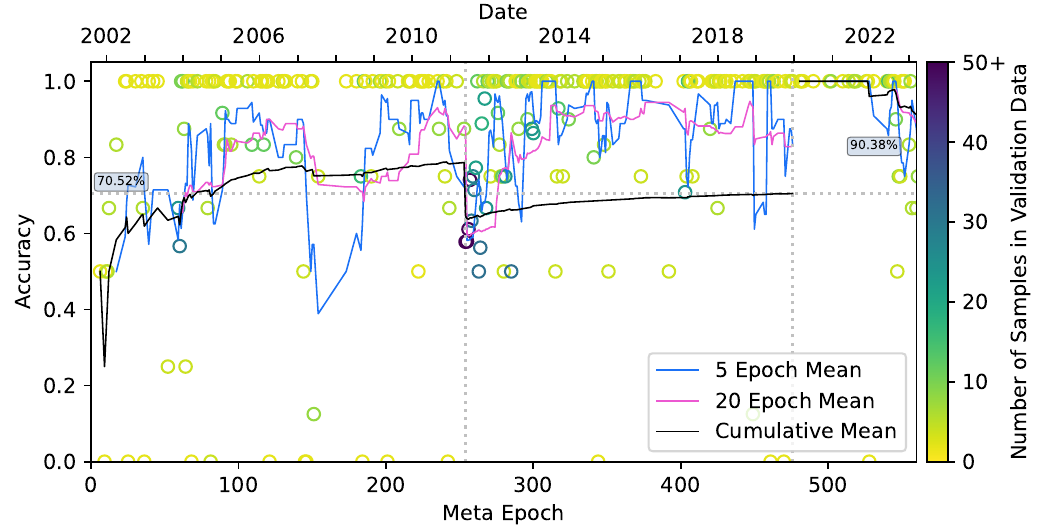}
    \caption{Classification Accuracy for meta epochs, analogous to Figure~\ref{fig:Res_Acc_ME}. For \MBest the validation accuracy during training, as well as the testing accuracy after 2019 is higher: $64.43\%$ for \MPerf and $70.52\%$ for \MBest during training and $81.41\%$ / $90.38\%$ during testing. Please note that these high values do not correspond to a realistic testing scenario, and should only be looked at as a internal metric.}
    \label{fig:Res_Acc_ME_boo}
\end{figure*}

\begin{figure*}
    \includegraphics[width=\textwidth]{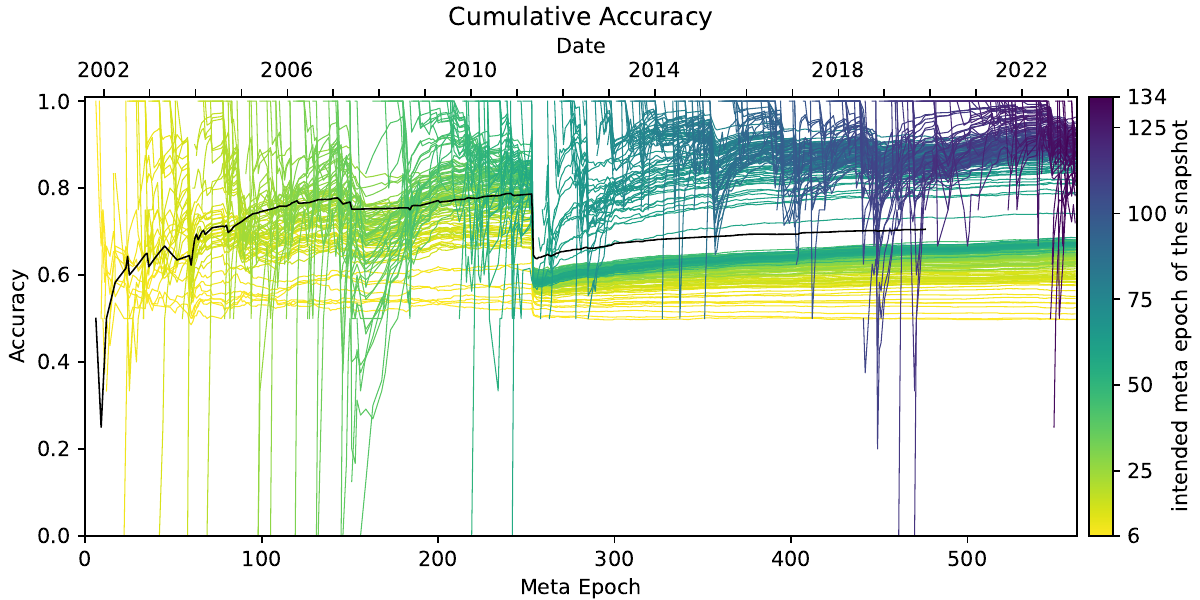}
    \caption{This figure shows the cumulative accuracy for each training snapshot after a meta epoch has finished in the colored lines, darker colors representing newer snapshots, analogous to Figure~\ref{fig:AccuracyDecay}. The black line is the overall cumulative accuracy when the always latest snapshot is used. It is noticable, that the \MBest case has very few poor performances after Tōhoku determined as dips to 0, while there were quite a few before that event. \MPerf on the other hand regularly has these happen, although not as often as with than \MBest before the Tōhoku Earthquake.}
    \label{fig:AccuracyDecay_boo}
\end{figure*}

\begin{figure*}
    \includegraphics[width=\textwidth]{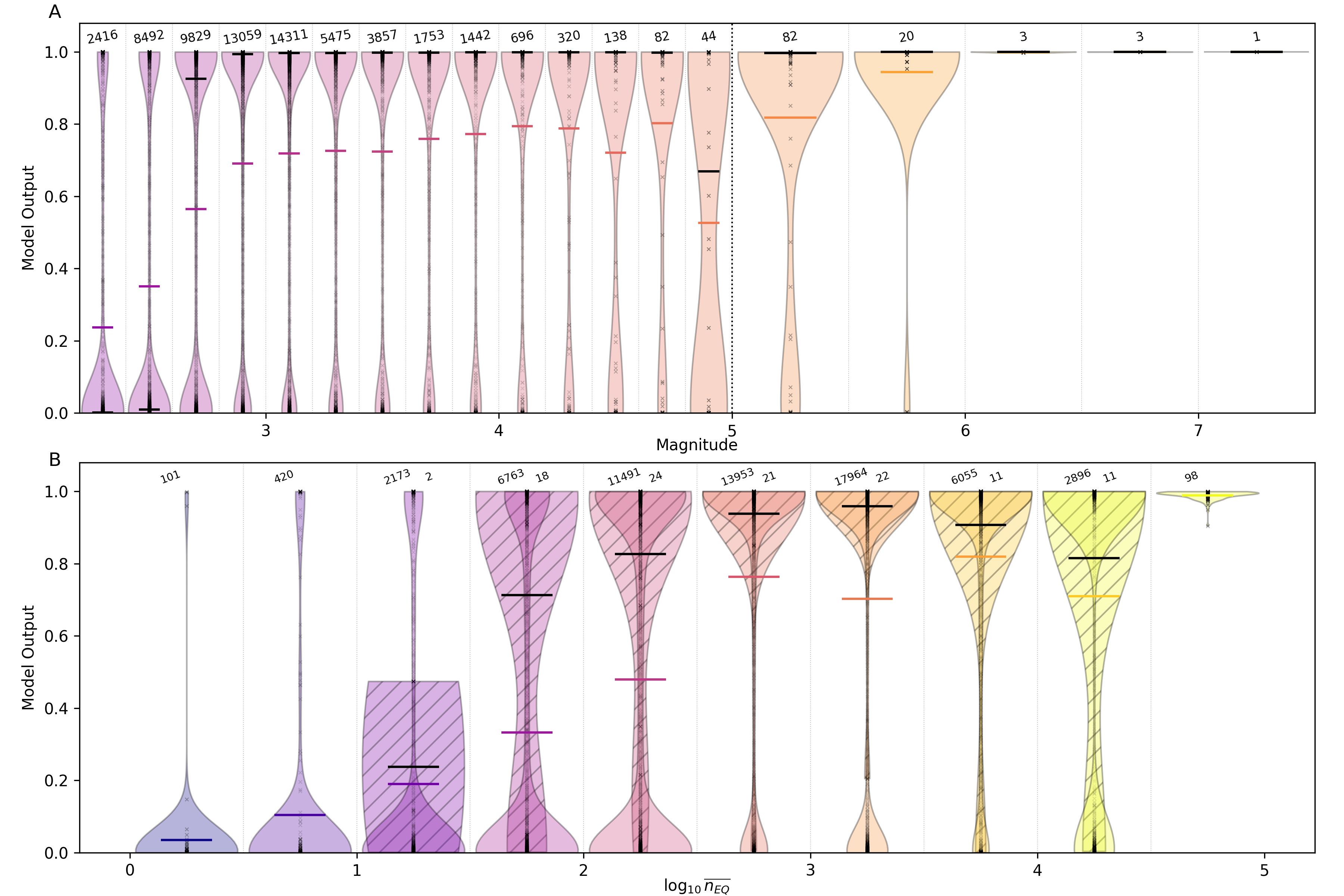}
    \caption{\MBest output as a function of earthquake magnitude and seismicity level, analogous to Figure~\ref{fig:Res_Hist_all}. The upper panel shows average model output (horizontal bars) and violin plots of model output with respect to event magnitude for the test set (post-2019). The number of events per magnitude bin is indicated above each violin. A clear increase in model output with event magnitude is observed. The lower panel displays model output as a function of the average number of earthquakes contributing to \bv calculation. Hatched violins and black vertical lines represent time series ending with events of magnitude \Mw $\geq$ 5 (\EQ), while colored violins and horizontal lines represent \nEQ\ time series (\Mw $<$ 5). Sample counts for both classes are shown above each bin \\(left: \nEQ, right: \EQ).}
    \label{fig:Res_Hist_all_boo}
\end{figure*}

\begin{figure}
    \includegraphics[width=\textwidth]{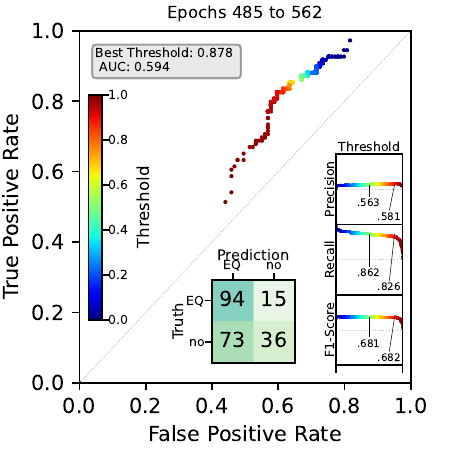}
    \caption{Receiver Operating Characteristic (ROC) curve and classification metrics for the best-performing model, evaluated on a balanced dataset from 2020 to 2022, analogous to Figure~\ref{fig:ROC_etc}. The optimal threshold derived from the ROC curve is 0.837; however, as shown in the inset plots (axes scaled 0–1), performance at the fixed threshold of 0.5 remains close to optimal. The confusion matrix summarizes classification results at the 0.5 threshold.}
    \label{fig:ROC_etc_boo}
\end{figure}

\begin{figure*}
\includegraphics[height=0.8\textheight]{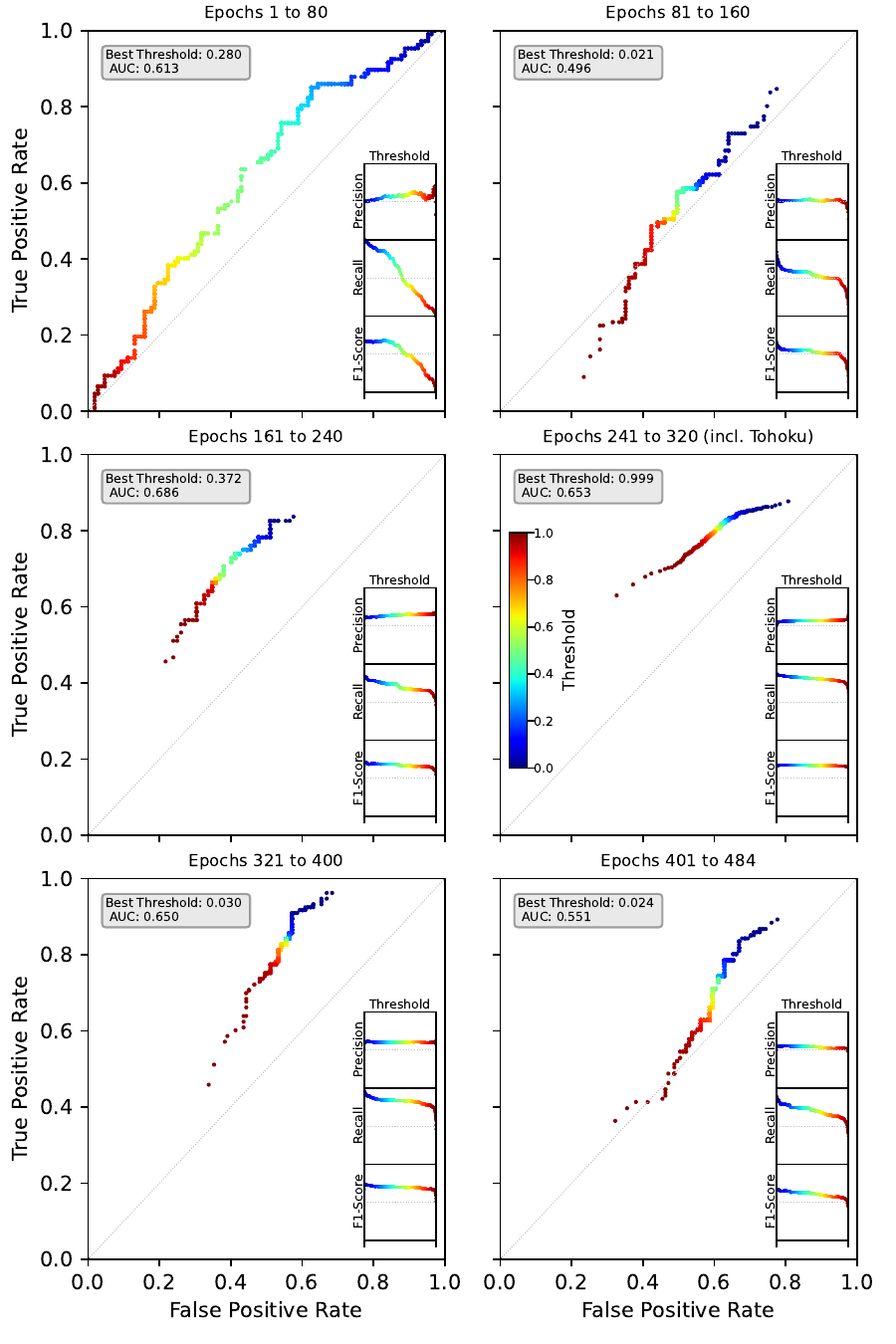}
\caption{Receiver Operating Characteristic (ROC) curves for different training intervals, analogous to Figure~\ref{fig:ROC_epochs}. The training period is divided into six approximately equal-length segments, chosen to match the duration of the interval used in Figure \ref{fig:ROC_etc}. A marked change is observed between intervals preceding and following the 2011 Tōhoku earthquake (occurring at meta epoch 255, near the beginning of the center-right panel). The influence of this major event is evident in both the panel containing the Tōhoku earthquake and the subsequent two panels, where substantially lower optimal thresholds are observed compared to the earlier intervals. Despite these shifts, a fixed threshold of 0.5 remains robust across all intervals.}
\label{fig:ROC_epochs_boo}
\end{figure*}

\end{document}